\documentclass{emulateapj}
\shorttitle{Star formation in IC~2574}
\shortauthors{Pasquali et al.}

\begin{document}

\title{The LBT Panoramic View on the Recent Star-Formation Activity in 
IC~2574}

\author{A. Pasquali, A. Leroy, H.-W. Rix, F. Walter, T. Herbst}
\affil{Max-Planck-Institut f\"ur Astronomie, K\"onigstuhl 17,
D-69117 Heidelberg, Germany}

\author{E. Giallongo}
\affil{INAF, Osservatorio Astronomico di Roma, via Frascati 33, I-00040 Monteporzio, 
Roma, Italy}

\author{R. Ragazzoni, A. Baruffolo}
\affil{INAF, Osservatorio Astronomico di Padova, Vicolo dell'Osservatorio 5, I-35122, 
Padova, Italy}

\author{R. Speziali}
\affil{INAF, Osservatorio Astronomico di Roma, via Frascati 33, I-00040 Monteporzio,
Roma, Italy}

\author{J. Hill}
\affil{Large Binocular Telesope Observatory, University of Arizona, 933 N. Cherry Ave., 
Tucson, AZ 85721-0065}

\author{G. Beccari}
\affil{Dipartimento di Astronomia, Universit\'a di Bologna, via Ranzani 1, 
I-40127 Bologna, Italy}

\author{N. Bouch\'e, P. Buschkamp}
\affil{Max-Planck-Institut f\"ur Extraterrestrische Physik, Giessenbachstrasse, D-85748 
Garching, Germany} 

\author{C. Kochanek}
\affil{Department of Astronomy, The Ohio State University, 140 W. 18th Ave., Columbus, OH 43210\\
Center for Cosmology and AstroParticle Physics, The Ohio State University, 191 W. Woodruff Ave.,
Columbus, OH 43210}

\author{E. Skillman}
\affil{Department of Astronomy, University of Minnesota, 116 Church St. SE Minneapolis, MN 55455}

\and

\author{J. Bechtold}
\affil{Steward Observatory, University of Arizona, 933 N. Cherry Ave., Tucson,
AZ 85721-0065}

\begin{abstract}
We present deep imaging of the star-forming dwarf galaxy IC~2574 in the M81 group taken 
with the Large Binocular Telescope in order
to study in detail the recent star-formation history of this galaxy and to constrain the 
stellar feedback on its HI gas. 
We identify the star-forming areas in the
galaxy by removing a smooth disk component from the optical images. We construct pixel-by-pixel 
maps of stellar age and stellar mass surface density in these regions 
by comparing their observed colors 
with simple stellar populations synthesized with STARBURST99. We find that  
an older burst occurred about 100 Myr ago within the inner 4 kpc and that a younger burst
happened in the last 10 Myr mostly at galactocentric radii between 4 and 8 kpc. 
We compare stellar ages and stellar mass surface densities with the HI column densities 
on subkiloparsec scales. No correlation is evident between star formation 
and the atomic H gas on local scales, suggesting that star formation in IC~2574 does not locally
expel or ionize a significant fraction of HI. 
Finally, we analyze the stellar populations residing in the known HI holes 
of IC~2574. Our results indicate
that, even at the remarkable photometric depth of the LBT data, there is no  clear one-to-one
association between the observed HI holes and the most recent bursts of star formation in
IC~2574. This extends earlier findings obtained, on this topic, for other dwarf
galaxies to significantly fainter optical flux levels. 
The stellar populations formed during the 
younger burst are usually located at the periphery of the HI holes and are seen to be 
younger than the holes dynamical age. The kinetic energy of the holes expansion is found
to be on average 10$\%$ of the total stellar energy released by the
stellar winds and supernova explosions of the young stellar populations within
the holes. With the help of control apertures distributed across the galaxy we estimate
that the kinetic energy stored in the HI gas in the form of its local velocity dispersion
is about 35$\%$ of the total stellar energy (and 20$\%$ for the HI non-circular motions), 
yet no HI hole is detected at the position of
these apertures. In order to prevent the HI hole formation by ionization, we estimate an 
escape fraction of ionizing photons of about 80$\%$. 
\end{abstract}


\section{Introduction}
It is common use to summarize with the term ``star formation'' how stars form from
the neutral gas reservoir of a galaxy, and how they evolve and affect the
ambient interstellar medium (ISM). Through their winds and supernova explosions,
stars release kinetic energy and freshly synthesized metals to the ISM,
with the possible effect of locally suppressing the formation of new stars
and triggering it on somewhat larger scales across their host galaxy (i.e. stellar
feedback). The global properties of disk galaxies have been extensively
studied by now, to find that the average star formation rate (SFR) of a galaxy
tightly correlates with its gas surface density (Schmidt 1959, 1963, 
Kennicutt 1998a,b). This important result alone, however, does not 
put strong constraints on the physics of star formation itself; it has 
been indeed shown that different mechanisms can produce this observed correlation 
(e.g. Silk 1997, Kennicutt 1998b, Elmegreen 2002). Moreover, our 
knowledge of star formation is mainly based on local disk galaxies,
chemically and dynamically different from local, dwarf irregulars which  
sometimes undergo intense starbursts (Heckman 1998). Because of their size and low
metallicity, dwarf galaxies are considered to be the local analogs to the
galaxy population at high redshift, where starbursts are observed to be
a rather recurrent phenomenon (Steidel et al. 1996, Pettini et al. 2001, 
Blain et al. 2002, Scott et al. 2002). Therefore, understanding the details 
of star formation in nearby galaxies of different Hubble types is crucial
for boot-strapping theories of galaxy formation and evolution. This is now made possible
by a new generation of instruments which provide high angular resolution
throughout the whole spectral range, and especially for the observations of
atomic H gas at 21 cm (HI). 
These new data allow us to spatially resolve the star-formation 
activity and the gas distribution within a galaxy, and to analyze their
correlation on small scales and across a wide range of local,
physical conditions. This approach has been recently applied by Kennicutt
et al. (2007) to M~51a and to a larger sample of galaxies by Bigiel et al. (2008)
and Leroy et al. (2008). 
\par  
In this paper, we apply a similar method  to the dwarf galaxy IC~2574 with the aim
of describing its star-formation activity (i.e. age and mass surface density
of its stellar populations) and comparing those properties with the interstellar
medium (i.e. HI) on a $\simeq$ 100 parsec scale. 
\par 
IC~2574 (aka UGC~5666, DDO~81) is a gas-rich dwarf, actively forming stars in
the M81 group. Miller \& Hodge (1994) observed it in the H$\alpha$ emission and
identified 289 HII regions, each about 50 pc in diameter. The largest complex
of star-forming regions is located in the North-East of the galaxy
(also known as a HI supergiant shell, Walter et al. 1998, Walter \& Brinks 1999). The 
authors estimated a global star-formation rate SFR $\simeq$ 0.08 M$_{\odot}$yr$^{-1}$ 
and concluded that IC~2574 is forming stars at roughly its average rate.   
Follow-up spectroscopy on some HII regions in the NE complex by
Tomita et al. (1998) showed that the H$\alpha$ velocity field is chaotic in all
regions except two (IC2574-I and IC2574-IV), which are characterized by a V-shaped 
velocity distribution. This V-shaped feature is usually representitative of stellar 
winds blowing out the interstellar medium. Interestingly enough, Drissen et al. (1993) 
found three candidate Wolf-Rayet (WR) stars in IC2574-I. WRs are evolved massive stars
known to develop stellar winds stronger than O stars of the same luminosity and to 
make up as much as 50$\%$ of the total mass and kinetic energy released by all massive
stars into the interstellar medium (cf. Leitherer, Robert \& Drissen  1992). Therefore, their
presence in IC2574-I can well explain the bulk motion observed in this HII region,
and also date the region to be a few Myrs old, given that the WR  phase is as short as
 2 - 5 Myr (Robert et al. 1993).

\begin{figure*}
\epsscale{.70}
\plotone{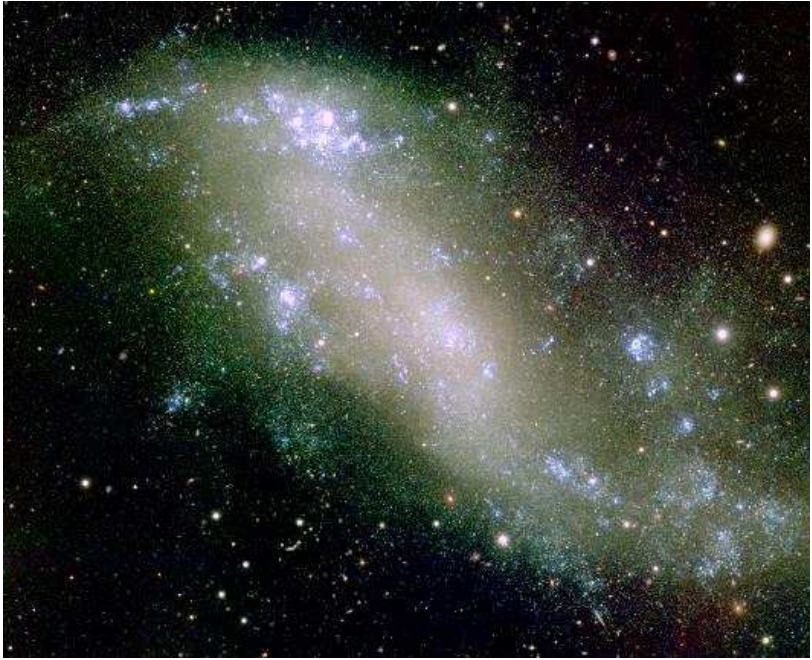}
\caption{Color composite of IC~2574 obtained with the LBT
LBC Blue camera in the Bessel $U$ (in blue), $B$ (in green) and $V$ (in red)
filters, during the Science Demonstration Time of February 2007.
North is up and East to the left. The size of the image is 14' $\times$ 11',
and corresponds to the central one fourth of the LBC field of view.}
\end{figure*}

The most intriguing property of IC2574 is, however, the spatial distribution of its
neutral interstellar medium. The high-resolution VLA observations of IC~2574 have
resulted in a very structured HI map, with 48 holes ranging
in size from $\sim$ 100 pc to $\sim$ 1000 pc. These features
appear to be expanding with a radial velocity of about 10 km~s$^{-1}$, which, in turn,
defines dynamical ages and kinetic energies consistent with being driven 
by stellar winds and supernova explosions (Walter \& Brinks 1999). In particular,
the comparison of the HI map with ultraviolet (UIT) and optical images has revealed that 
the HII regions
located in the NE complex are distributed along the rim of the HI supergiant shell,  
and the shell itself is devoided of neutral and ionized gas, possibly filled with
hot X-ray gas (Walter et al. 1998). Stewart \& Walter (2000)
identified a star cluster within the supergiant shell with no associated H$\alpha$
emission  and with an age (11 Myr) comparable to the shell dynamical age of 14 Myr.
They dated the H$\alpha$-emitting rim of the shell to be about 3 Myr old. The 
authors thus concluded that the HI supergiant shell was created by this star
cluster and the shell expansion triggered the star formation activity along its rim.   
As pointed out by Walter \& Brinks (1999), the occurrence of H$\alpha$ emission 
along the rim of a HI feature is common to all the HI holes found in
IC~2574, even if no remnant central star cluster has been identified so far. The question 
is then whether these holes are the result of stellar
feedback as in the case of the supergiant shell  
or are produced by different mechanisms. The data collected
so far for the Magellanic Clouds (Kim et al. 1999, Hatzidimitriou et al. 2005),
M31 and M33 (cf. van der Hulst 1996) and Holmberg II (Rhode et al. 1999)   
offer very little support for a one-to-one correlation between HI holes and OB
associations, although the kinetic energy of the hole expansion is consistent with
the expansion being triggered by the total energy released by a 
single stellar population. One possible explanation that has been put forward is that 
at least some HI holes (especially those 
with no indication of recent star formation) may have been produced by mechanisms 
unrelated to young
stars. For example, Loeb \& Perna (1998) and Efremov, Elmegreen 
\& Hodge (1998) suggest they arise from  Gamma-Ray bursts. 
HI holes may also be produced by the collision of high-velocity clouds
of neutral gas with the galactic disk (cf. Tenorio-Tagle 1980, 1981), or, more simply,
by the intrinsic turbulence of the interstellar medium (Elmegreen \& Efremov 1999,
Dib, Bell \& Burkert 2006). Rhode et al. (1999) proposed the intergalactic UV radiation 
field as an additional 
mechanism to create holes in the HI gas by keeping it ionized (at least in the 
outskirts of the galactic disk where the HI volume density is lower), while Bureau \&
Carignan (2002) suggested that the HI holes detected in Holmberg II may have formed
by ram pressure from the intragroup medium.
\par 
One could also imagine that the lack of a one-to-one correlation between HI holes and OB
associations may be due to the limited photometric depth of the data. For example,
Rhode et al. (1999) reached a limiting magnitude $B \simeq$ 23 mag in Holmberg II
and may have failed to detect  fainter young stars (i.e. late B spectral types). 
We have taken advantage of the high sensitivity and wide field of view of the
LBC Blue camera mounted on the Large Binocular Telescope to perform deep optical
imaging, to study in detail the
stellar populations in IC~2574  and to investigate the possible {\it stellar origin}
of the HI holes in this galaxy. 
Our aim is to identify low-luminosity star clusters within these features, 
to establish how their stellar content may be different from any other position 
across the galaxy, and whether the energy budget of the young stars in the HI
holes can account for the kinetic energy of these structures. 
This study allows us also to derive the star-formation history of
IC~2574 locally,
at different galactocentric radii and in correspondence with different HI
column densities. The observations and the data reduction are described  
in Sect. 2. The high-resolution, optical morphology of IC~2574 is presented
in Sect. 3, while the stellar ages and masses resulting from our dating
technique are discussed in Sect. 4. We describe the stellar populations associated
with the HI holes in Sect. 5. and draw our conclusions in Sect. 6. 

\section{Observations}
The observations presented here were carried out with the Large Binocular Telescope
(LBT) located on Mount Graham, Arizona, (Hill et al. 2006). IC~2574
was observed as part of the Science Demonstration Time for the LBT
blue camera (LBC-Blue, Ragazzoni et al. 2006, Giallongo et al. 2008),
a wide-field imager with a 23$' \times$ 23$'$ field of view and an angular
scale of 0.23 arcsec/pixel. LBC-Blue is optimized for high throughput
across the spectral range from 3200 to 5000 \AA\/ and has 4 CCD detectors of
2048 $\times$ 4608 pixels each. Its filter set includes the Bessel $U$, $B$ 
and $V$ filters, plus the $g$ and $r$ bands. At the distance of IC~2574, 
4.02 Mpc (Walter et al. 2007), one LBC-Blue pixel corresponds to about 4.4 pc.

IC~2574 was imaged with the LBT/LBC-Blue on February 17, 19, 21 and 22,
2007, using a circular dithering pattern (40 arcsec in radius) with 4 different
pointings in order to cover also the 3 intra-CCD gaps (about 16 arcsec each) in the
LBC-Blue field of view. The total exposure time was 70 min 
(14 $\times$ 300 s) in the $U$ band, 35 min (7 $\times$ 300 s) and 15 min 
(3 $\times$ 300 s) in the $B$ and $V$ bands, respectively, with an average seeing of
$\sim$ 1.0$''$. The individual images were reduced with standard IRAF\footnote{IRAF 
is distributed by  the National Optical  Astronomy Observatories, 
which are  operated by the  Asssociation of  Universities for  Research in
Astronomy,  Inc.,  under  cooperative  agreement with  the  National Science  Foundation}
routines. They were corrected for bias and flat-field, and then aligned to the same 
pixel grid using GEOMAP and GEOTRAN with an accuracy of better than 0.2 pixels. After 
background subtraction, the images obtained for the same filter were weighted by their
seeing (weight = $\sigma^2_{image}$/$\sigma^2_{best}$, where $\sigma^2_{best}$ is the
best seeing measured for each filter dataset) and averaged. The astrometric solution
of these combined images was updated using the package Astrometry.net (http://astrometry.net,
Lang et al. 2008). We applied the flux calibration derived by the LBC-Blue
team during commissioning to the combined images, and checked the consistency of these 
color-transformations on standard stars imaged at the same epoch as IC~2574. The systematic error 
of the flux calibration turns out to be $\sim$0.03 mag in each band. The 
combined images reach a limiting surface brightness of 26 mag~arcsec$^{-2}$ (in the VEGAMAG system)
at a S/N ratio of 5 in each filter. 

In what follows, the LBT optical data collected for IC~2574 are used together with
the observations in atomic H gas (HI at 21 cm) performed by Walter \& Brinks (1999)
with the VLA in its B, C and D configurations. The HI observations were carried out with 
a 1.56 MHz bandwidth centered at a heliocentric velocity of 38 km~s$^{-1}$, for
a velocity resolution of 2.6 km~s$^{-1}$ and an angular resolution of about 7 arcsec.
The HI map defines a maximum rotation velocity of 70 km~s$^{-1}$ at a galactocentric
distance of 8 kpc (Oh et al. 2008) and an average velocity dispersion of 7 km~s$^{-1}$
(Walter \& Brinks 1999). 

\begin{deluxetable}{lccc}
\tablecaption{Exponential disk parameters}
\tablewidth{0pt}
\tablehead{
\colhead{} & \colhead{$U$}  & \colhead{$B$} & \colhead{$V$}\\
}
\startdata
$\mu_0$ (mag~arcsec$^{-2}$) & 23.72 & 23.94 & 23.47\\
$R_o$ (kpc)                 & 4.1 & 3.6 & 3.4\\
$m_T$ (mag)                 & 10.67 & 11.13 & 10.70 \\
$m_T^0$  (mag)              & 10.49 & 10.98 & 10.59 \\
\enddata
\end{deluxetable}

\section{The optical morphology of IC~2574} 
The color-composite of IC~2574 obtained with the LBT/LBC Blue camera during the
Science Demonstration Time of February 2007 is shown in Figure 1, with the $U$, $B$ and
$V$ images color-coded in blue, green and red respectively. The size of the
color-composite in Figure 1 is 14$' \times$ 11$'$ and covers only a quarter of the full field
of view of the camera. The combination
of high sensitivity with wide field of view of the camera reveals
a wealth of features: the ``finger'', a knotty tidal stream (oriented East-West),
the Northeast complex of HII regions (the HI supergiant shell), and  
the extended tail of star formation in the South-West.
Faint plumes of stars departing from the eastern and
western sides of IC~2574 are also visible. The star-formation activity in IC~2574
appears to mostly occur in the outskirts of the galaxy, in a ring-like
substructure. This is reminiscent of the optical morphology of the  
Large Magellanic Cloud, which, like IC~2574, is a barred late-type galaxy. 
\par
From the isophote corresponding to a limiting surface brightness of 26 
mag~arcsec$^{-2}$ we measure an overall extend for IC~2574 of about 16 kpc
$\times$ 7 kpc, which includes the East-West tidal 
stream and the South-West tail. 
The ellipse that best fits the $\mu$ = 26 isophote is 
characterized by an axis ratio
$q$ = $b/a$ = 0.47 and a position angle PA  = 55$^o$. For a disk of intrinsic flattening 
of $q_0$ = 0.12 (cf. Martimbeau et al. 1994) this implies an inclination
$i$ = 63$^o$. With these orientation parameters, we compute
the surface brightness profile for concentric elliptical annuli of fixed
$q$ and PA in each 
band. These profiles are plotted in Figure 2, where the surface brightness is
corrected for inclination (assuming no dust) and Galactic extinction 
(E($B$-$V$) = 0.036 mag, Schlegel et al. 1998) using the extinction law of 
Cardelli et al. (1989). These profiles nicely follow an exponential disk
out to $R \simeq$ 4 kpc, with an excess of emission at $R \simeq$ 5 and 7 kpc 
(see Figure 1). The parameters of the best-fitting
exponential disk are listed in Table 1 for each filter, where the central 
surface brightness ($\mu_0$) is corrected for inclination and the total observed 
magnitude ($m_T$) is corrected for Galactic extinction ($m^0_T$). The mean 
scale length computed from the LBT data is $R_o \simeq$ 3.7 kpc and agrees well with the
scale length (3.2 kpc) obtained by Martimbeau et al. (1994, adopting the same distance). 
The inner disk (at $R <$ 4 kpc) is characterized by a mean $(B-V) \simeq$
0.44 mag and a mean $(U-B) \simeq$ -0.3 mag (both corrected for Galactic extinction). 
In a model of an instantaneous starburst (i.e. a single stellar population, SSP)
computed with STARBURST99 (Leitherer et al.
1999) for a metallicity similar to the Small Magellanic Cloud (SMC), these colors translate 
to an age of $\sim$ 1 Gyr and 120 Myr, respectively.

Figure 3 compares the $\mu$ = 26 $U$-band isophotal contour (thick white line) to the HI 
distribution to illustrate the spatial relationship between the young stars and the 
resevoir of atomic gas. 
The grey scale becomes darker as the column density of the HI gas 
increases from 5 $\times$ 10$^{20}$ to about 5 $\times$ 10$^{21}$ cm$^{-2}$ 
(not corrected for inclination). While the HI and $U$-band light distributions share
nearly the same major axis, their minor axes are quite different, with the minor
axis being smaller in the optical. This could be due to differences in the scale
height of the $U$-band light and HI distributions. 

\begin{figure*}
\epsscale{.60}
\plotone{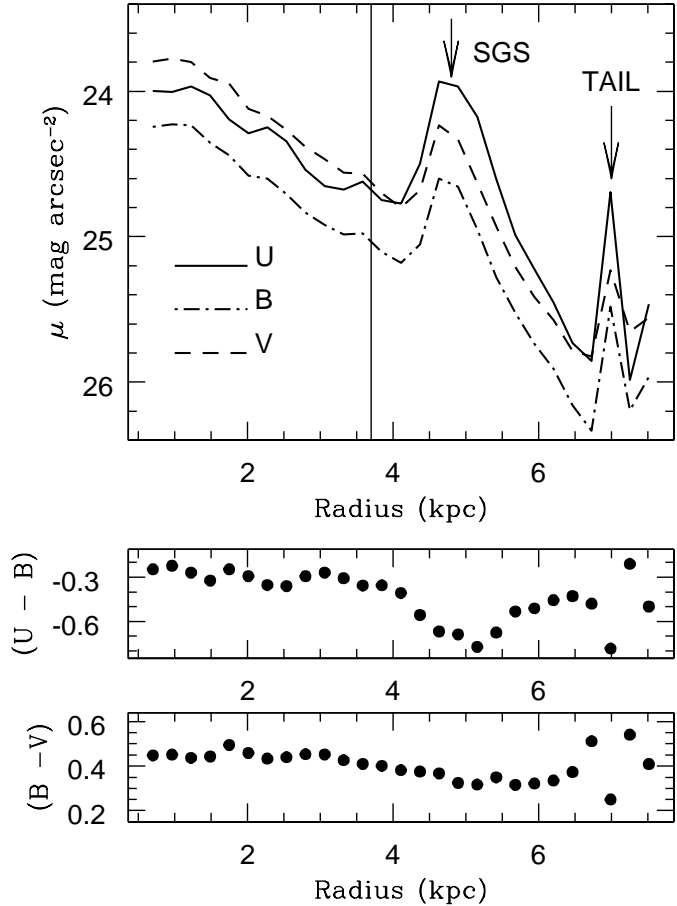}
\caption{The surface brightness profile computed in the $U$, $B$ and $V$ bands,
together with the radial profile of the $(U-B)$ and $(B-V)$ colors. The surface
brightness is corrected for inclination and Galactic extinction. The colors
are also corrected for Galactic extinction. The vertical solid line indicates
the scale length of IC~2574, $R_o$ = 3.7 kpc.}
\end{figure*}

\begin{figure}
\epsscale{.80}
\plotone{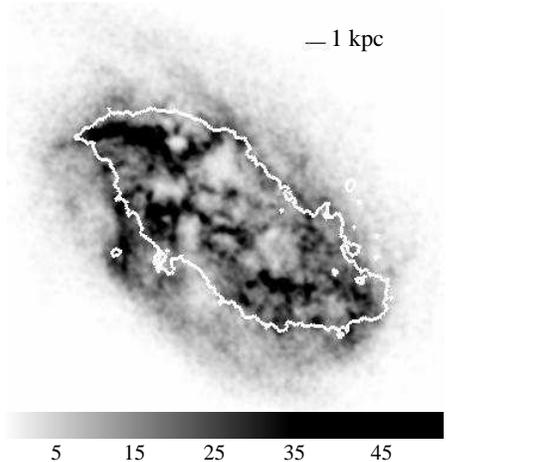}
\caption{The isophotal contour of IC~2574 in the $U$ band corresponding to $\mu_U$
= 26 mag~arcsec$^{-2}$ (thick white line) superposed on the map of the HI column 
density
in grey (from Walter \& Brinks 1999). The grey scale becomes darker as the column
density increases from 5 $\times$ 10$^{20}$ cm$^{-2}$ to about 50 $\times$ 10$^{20}$
cm$^{-2}$.}
\end{figure}

\section{Stellar populations} 
In order to derive the star-formation history of IC~2574 over the last few 10$^8$ yr
from the observed photometry, we need to disentangle the colors of star-forming regions from those
of the underlying smooth disk, which includes light from older episodes of
star formation (see Sect. 3). This is analogous to subtracting the local background in
stellar photometry. We first identify the star forming regions as those 
bluer than $(U - B)$ = -0.3 mag and with a S/N ratio per pixel larger 
than 5 (in all filters). These regions are likely to contain mostly light from young
stars (younger than $\simeq$ 100 Myr for a SSP synthesized with STARBURST99 assuming 
an SMC-like metallicity). We mask these regions and then fit the remaining light. 
We fix the galaxy center, ellipticity (1 - $q$ = 0.53), and position
angle (55$^o$) and fit the $U$-band image together with the mask frame using
the IRAF routine ELLIPSE. This yields an estimate of the diffuse 
light across the galaxy and we subtract this model disk component from the
original $U$ image.
The residual image should contain only young stars and we use
it in the subsequent analysis. Figure 4 shows,
as an example of the whole procedure, the original $U$ image (top panel), the model
disk component with the masked regions overlaid (middle panel) and the image of the
residuals (bottom panel) determined for the $U$ band. This same procedure was applied
to the $B$ and $V$ images after convolving them to have the same 
PSF as the $U$ band. Each band was fitted with ELLIPSE separately, but with the same
mask and the same fixed values for the galaxy center, position angle and ellipticity. 
\par
The resulting model disk component is characterized by a scale length of 2.7 kpc in all
three filters, about 30$\%$ smaller than the average $R_o$ computed in Sect. 3
from the original images (disk $+$ star-forming regions). Its mean colors are
$(B -V) \simeq$ 0.5 mag and $(U - B) \simeq$ -0.2, which correspond to an age of
about 1 Gyr and 200 Myr, respectively, according to a SSP synthesized with STARBURST99
for an SMC-like metallicity. These values are consistent with the ages derived from the 
color profiles for $R <$ 4 kpc; therefore, subtracting the model disk component is equivalent
to removing the light from the diffuse stellar populations older than $\sim$100 Myr. 
The model disk component makes up about 27$\%$, 36$\%$ 
and 42$\%$ of the total light of the galaxy in the $U$, $B$ and $V$ filters respectively.

The observed $(U-B)$ and $(B-V)$ colors of the pixels in the residual images 
are plotted in Figure 5 as a function 
of distance from the galaxy center. These colors have been corrected only for Galactic 
extinction. 
%(E($B$-$V$) = 0.036 mag, Schlegel et al. 1998) using the extinction law of Cardelli 
%et al. (1989). 
The grey scale of the color distributions is based on a grid of bi-dimensional
bins, 0.4 kpc $\times$ 0.2 mag in size in $(U-B)$, and 0.4 kpc $\times$ 0.1 mag in size 
in $(B-V)$. We count the number of pixels falling in each bin and scale the grey shades 
by the maximum, so that the grey colors grow darker as the number of pixels increases. 
The right-hand side $y$-axis in both plots qualitatively links colors to stellar ages of
a SSP synthesized with STARBURST99 for an SMC-like metallicity (in units of 10$^6$ yr). 
We notice that the residuals within 2 kpc from the
galaxy center are relatively redder and older than those at larger distances, while there is
a significant blue (few Myr old) stellar population at galactocentric
radii larger than 3 - 4 kpc.

In what follows, we will restrict our analysis to the pixels in the residual images.
Our aim is to estimate their stellar age and stellar mass surface density (Sect. 4.1), with which 
we will reconstruct the star-formation history in IC~2574 in the last $\sim$10$^8$ yr 
(Sect. 4.2). We will compare the star-formation activity of this galaxy with its HI
properties to assess the applicability of the Schmidt law on local and global scales 
(Sect. 4.3). Finally, we will investigate the possible stellar origin of the HI holes 
detected in IC~2574 to examine the effects of stellar feedback 
on the interstellar medium (Sect. 5).  

\begin{figure}
\epsscale{.70}
\plotone{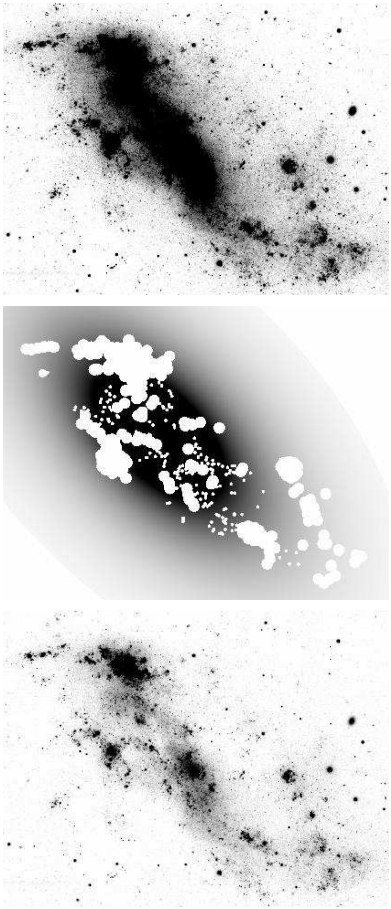}
\caption{An example of how the smooth disk component is subtracted from the original
images. The top panel shows the initial image, in this case the $U$ frame, the middle 
panel shows the model disk component obtained with ELLIPSE after having masked those 
regions in IC~2574 bluer than $(U - B)$ = -0.3 mag and with a S/N ratio larger than 5 
in all filters (white apertures). Finally, the bottom panel shows the residuals after 
subtracting this model from the original $U$ image. Flux increases as the grey shade 
becomes darker. All images are oriented so that North is up and East is to the left;
their size is 15 kpc $\times$ 11 kpc.}
\end{figure}

\begin{figure*}
\epsscale{1.00}
\plotone{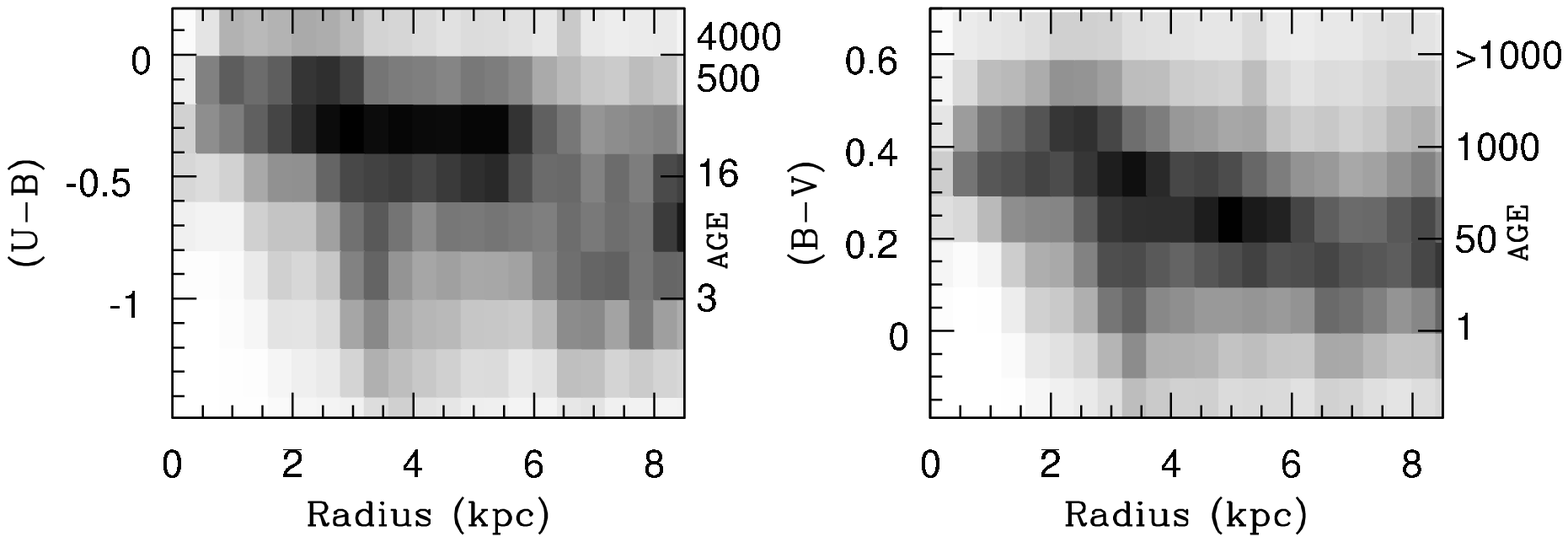}
\caption{The radial distribution of observed $(U-B)$ and $(B-V)$ colors in the
residual images (after subtracting the smooth disk model), corrected
only for Galactic extinction. The grey scale indicates the number of pixels in each
bi-dimensional bin [0.4 kpc $\times$ 0.2 (0.1) mag for $(U-B)$ and $(B-V)$], so that
it gets darker as the number of pixels increases. The right-hand side $y$ axis shows,
in both plots, the stellar ages corresponding to the colors on the left-hand side $y$
axis. These stellar ages are in units of 10$^6$ yr and refer to a SSP computed with
STARBURST99 for an SMC-like metallicity.}
\end{figure*}

\subsection{Age and mass estimates}
In order to assign an age and a mass density for each pixel in the residual images, we 
compare the observed photometry to synthetic stellar populations 
convolved with a specific reddening law. We run STARBURST99 to compute the colors of an
instantaneous starburst (SSP) with a total mass of 10$^6$
M$_{\odot}$ as a function of its age. For this purpose, we adopt the default parameters
of STARBURST99 except for the metallicity, which we choose to be SMC-like following
Stewart \& Walter (2000). We assume a multi-power law for the initial mass function (IMF)
to approximate a Kroupa (2001) IMF, with an exponent of 1.3 for stars in the range 0.1
M$_{\odot}$ - 0.5 M$_{\odot}$, and 2.3 for the interval 0.5 M$_{\odot}$ - 100 M$_{\odot}$.
Only stars with masses between 8 M$_{\odot}$ and 120 M$_{\odot}$ produce
supernovae. Stellar winds and mass loss are computed with the theoretical model which
solves for the radiative transfer and the momentum equation in the stellar wind (cf.
Leitherer, Robert \& Drissen 1992). Finally, we use the Geneva evolutionary tracks with 
high mass loss; as discussed by Meynet et al. (1994), these tracks better reproduce the
low luminosity observed for some Wolf-Rayet (WR) stars, the surface chemical composition of
WC and WO stars, and the ratio of blue to red supergiants in the star clusters in the
Magellanic Clouds. The choice of an instantaneous starburst may be an over-simplified
approach to the study of a star-formation history that is expected to be more complicated
in reality. On the other hand, our analysis is restricted only to star forming regions younger
than $\sim$10$^8$ yr and this short time interval can be represented with an instantaneous 
starburst reasonably well.

We select all pixels in the residual images with a S/N ratio larger than
5 in all filters and with a $U$-band flux 5$\sigma$ above the background level.
We compare their colors (corrected for Galactic extinction) with the synthetic colors of 
STARBURST99 using the
technique described in Pasquali et al. (2003). Briefly, the synthetic colors of a single
stellar population are reddened according to a pre-selected extinction law by
a quantity E($B$-$V$)$_i$ which is a free parameter varying across a pre-defined interval.
This parameter represents the dust obscuration within the galaxy itself, and we refer to
it as intrinsic reddening.
For a given E($B$-$V$)$_i$, a $\chi^2$ is calculated as the sum of the differences between the 
observed and reddened synthetic colors, normalized by the observational errors. Its minimum 
value, $\chi^2_{min}$, establishes the best-fitting age, intrinsic reddening and reddened
synthetic magnitudes in the $U$, $B$ and $V$ bands. The stellar mass enclosed in each pixel
is then determined via the scaling relation:
\par
M$_{pix}$ = 10$^{-0.4(m_{obs}-m_{syn})} \times$ 10$^6$ M$_{\odot}$ 
\par\noindent
We compute M$_{pix}$ in $U$, $B$ and $V$, and take the mean of these three
values.

Although at mid-IR wavelengths the intrinsic reddening of IC~2574 is negligible 
(cf. Cannon et al. 2005,
Dale et al. 2007), at optical wavelengths it appears to be more prominent. Miller \& Hodge 
(1996) performed optical spectroscopy of the North-East complex of HII regions and measured
an intrinsic reddening of the gas, E($B$-$V$)$_i^{gas}$, between 0.3 and 0.5 mag, which
translates into a stellar E($B$-$V$)$_i$ of 0.13 - 0.22 mag (cf. Calzetti 2001). More recently,
Moustakas et al. (2007) derived E($B$-$V$)$_i^{gas} \simeq$ 0.15 mag over an aperture 10
arcsec in radius at the center of IC~2574 which implies even lower stellar E($B$-$V$)$_i$. 
Therefore, we decide to restrict the range of values
available for E($B$-$V$)$_i$ in our fitting procedure between 0.0 and 0.13 mag and to apply 
reddening with the extinction law derived for the SMC by Bouchet et al. (1985). We also use
our dating technique in the assumption of a reddening-free environment (imposing E($B$-$V$)$_i$ 
= 0.0 everywhere in the galaxy) to check for systematic effects on the output ages and
mass densities due to our specific treatment of dust obscuration. 

In the dating technique of Pasquali et al. (2003), the errors on each best-fitting parameter are 
given by the minimum and maximum value of that parameter among all the fits  which realize  
$\chi^2 <$ 1.5$\chi^2_{min}$ (corresponding to about 2$\sigma$). In the case of IC~2574 we 
compute an average uncertainty of a factor of 2 on age, reddening and mass.

\begin{figure*}
\epsscale{0.80}
\plotone{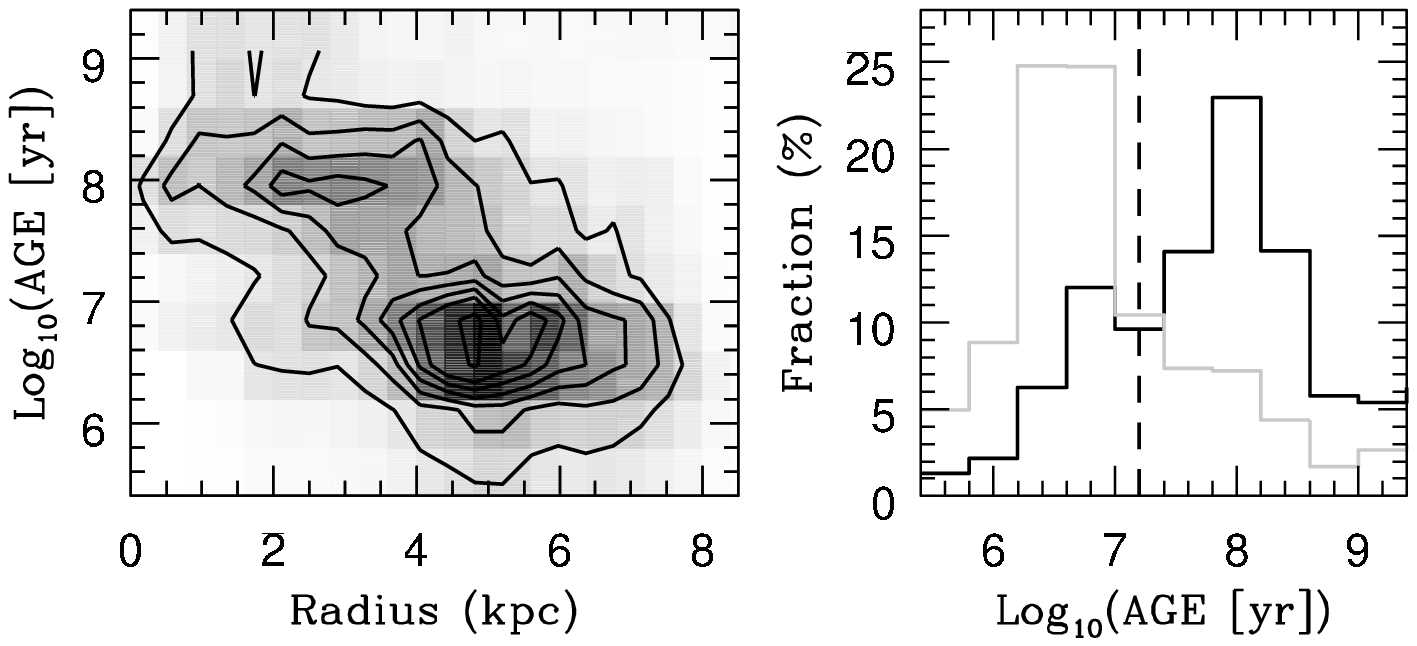}
\caption{The luminosity-weighted age distribution of the stellar populations across
IC~2574, obtained from the residual images after correction for reddening. The left hand-side
panel shows the distribution of pixel stellar age as a function of pixel galactocentric
distance. The grey scale indicates the number of pixels in each 2D bin, 0.4 dex $\times$ 0.4
kpc in size. The grey shade becomes darker with increasing  number of pixels. The black solid
lines are simply the contours drawn per constant number of pixels, with the following levels:
0.5, 1, 1.5, 2, 2.5, 3, 4 and 5 $\times$ 10$^4$. The right hand-side panel plots the
age histogram for all pixels with $R <$ 4 kpc (black line) and $R \geq$ 4 kpc
(grey line). Both histograms are normalized by the total number of pixels within and outside
4 kpc, respectively. The black dashed line corresponds to Log$_{10}$(AGE) = 7.2 (16 Myr),
the value we use to separate the younger from the older burst.}
\end{figure*}

\begin{figure*}
\epsscale{0.80}
\plotone{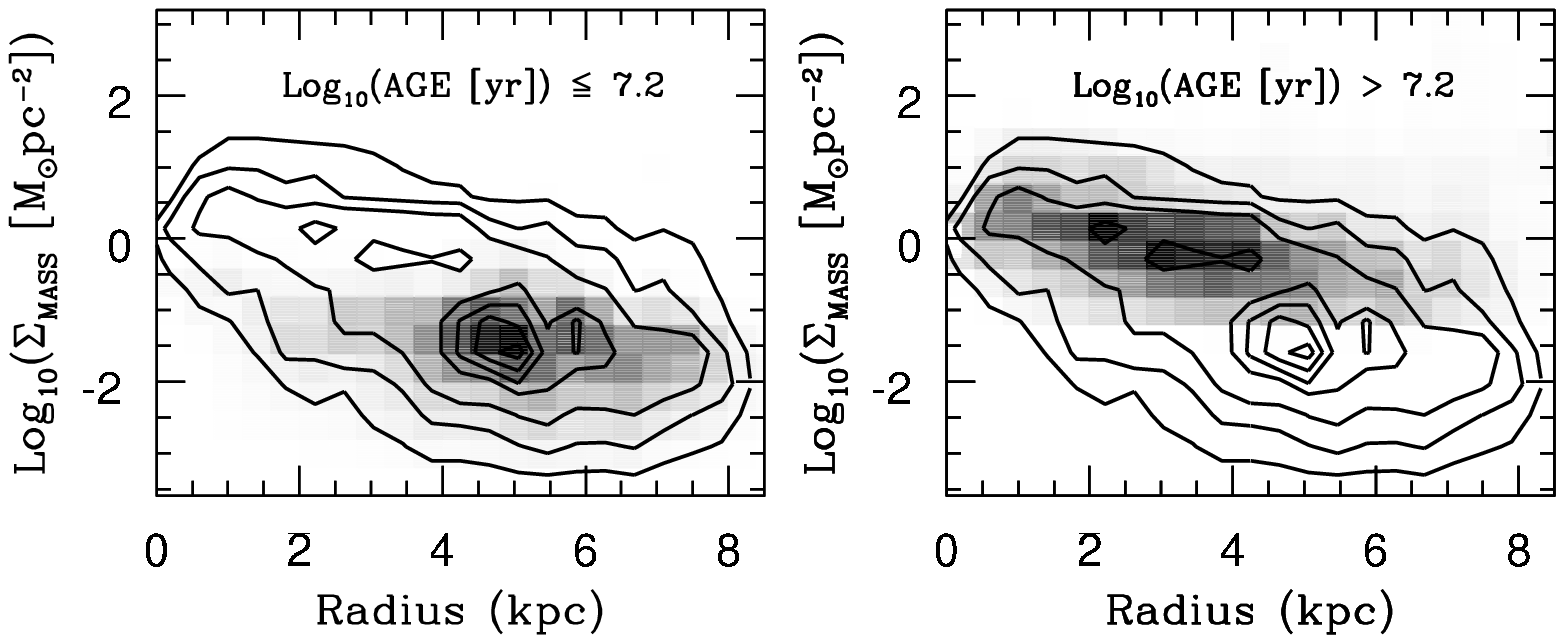}
\caption{The stellar mass surface densities distribution across IC~2574 obtained after
correction for reddening. The left hand-side panel shows the distribution of the
mass surface density produced by the younger burst, while the right hand-side panel
shows the radial variation of the mass density
obtained for the older burst. Once again, the grey scale indicates the number of pixels
in each 2D bin, 0.4 dex $\times$ 0.4 kpc in size. Its shades
grow darker as the number of pixels increases. The black contours trace bins of equal number
of pixels, and are drawn for the following levels: 0.2, 0.5, 1, 2, 2.5, 3, 3.5
and 4 $\times$ 10$^4$ pixels. They have been computed for all the pixels with no
age selection. All stellar mass surface densities have been
corrected for inclination.}
\end{figure*}

\subsection{The spatial distribution of stellar ages and mass densities}
The distribution of stellar ages (obtained by correcting for intrinsic reddening) 
across the residual images
is shown in Figure 6. The left hand-side panel plots the distribution of pixel
stellar ages as a function of pixel galactocentric distance with the help of a grey scale 
which indicates the number of pixels in each 2D bin, 0.4 dex in age $\times$ 0.4 kpc in radius. 
The grey shade gets darker with increasing numbers of pixels. The contours (tracing equal
numbers of pixels across the bins) highlight two major epochs of star formation in IC~2574.
The older burst occurred about 100 Myr within $R <$ 4 kpc of the center, while the
second event took place between 10 and 1 Myr ago mostly at $R \geq $ 
4 kpc. Star formation proceeded in between the two bursts, but with a somewhat lower
rate. Globally, the stellar ages decrease with increasing distance 
from the galaxy center. 
 
The age histograms for all pixels at a galactocentric distance smaller and larger
than 4 kpc are traced in the right hand-side panel of Figure 6, with a solid black and grey
line respectively. Both histograms are normalized by the total number of pixels from which they
are derived. If we assume a value of 16 Myr (Log$_{10}$(AGE) = 7.2) to discriminate 
between the younger and older burst, we can estimate the spatial coverage of the two 
bursts in the inner and outer disks of IC~2574. For galactocentric distances smaller than 4 kpc, 
the older burst has occurred over 73$\%$ of the pixels while the younger one has involved 
only 27$\%$ of the area. In the outer regions, instead, the younger burst has lit up 70$\%$ 
of the pixels, and the older burst took place only in 30$\%$ of the area.  
\par
Repeating the same analysis but switching off the reddening correction confirms these trends. 
The younger and older bursts are still prominent and well defined, 
and they turn out to have similar timescales as determined when reddening is applied.
Only the spatial coverage of the two bursts varies, and this only by 5$\%$ at most. 

The distribution of the pixel, stellar mass surface densities (obtained including the correction for
intrinsic reddening) as a function of pixel galactocentric distance is presented in Figure 7 for 
the younger burst (left hand-side panel) and the older burst (right hand-side panel). Here, 
the grey scale indicates the number of pixels in each 2D bin, 0.4 dex $\times$ 0.4 kpc in size. 
The contours, which trace equal numbers of pixels across the bins, are computed for all the
pixels with no age selection, and are the same in both panels. All stellar mass surface densities 
are corrected for inclination. We can see 
that the older burst involves stellar mass surface densities larger than 
$\sim$0.2 M$_{\odot}$pc$^{-2}$ and 
has a characteristic stellar mass surface density of about 1 M$_{\odot}$pc$^{-2}$. 
The younger burst is 
instead associated with stellar mass surface densities lower than 0.2 M$_{\odot}$pc$^{-2}$ with 
a characteristic 
value of about 0.04 M$_{\odot}$pc$^{-2}$. The stellar mass surface density contours trace  
a radial profile which follows an exponential disk with a scale length of about 2.7
kpc and a central mass density of $\sim$5 M$_{\odot}$pc$^{-2}$. 
\par
Correcting for reddening yields lower ages, higher stellar luminosities and hence 
larger stellar masses than assuming no extinction. With 
%our adopted approach (E($B$-$V$)$_i$ fit under 
the constraint that E($B$-$V$)$_i$ is $<$ 0.13 mag
see Sect. 4.1), we obtain mass surface densities that are up to 40$\%$ larger
than those assuming E($B$-$V$)$_i$ = 0.0 everywhere. 
The assumption of E($B$-$V$)$_i$ = 0.0 everywhere in the
galaxy results in a lower characteristic stellar mass surface density, of about 
0.6 M$_{\odot}$pc$^{-2}$ and
0.03 M$_{\odot}$pc$^{-2}$ for the older and younger burst respectively. It also gives a slightly
larger scale length of 3 kpc and a central, stellar mass surface density of $\sim$ 6 
M$_{\odot}$pc$^{-2}$.    

Independent of the reddening, there is a relationship between age and mass surface density 
in our results, in the sense that older stellar ages are associated with higher, stellar 
mass surface densities.  This is shown
in more detail in Figure 8, where the stellar mass surface density (corrected for inclination) 
is plotted 
as a function of age. The grey scale is based on the number of pixels within each
2D bin, 0.4 dex $\times$ 0.4 kpc in size. 
The lower edge of the distribution is a selection effect created by requiring a $U$-band
detection in order to carry out the analysis.  Since older stars emit less $U$-band light, 
there must be more of them to reach the same surface brightness limit.  The solid line in 
Figure 8 shows that the lower edge of the relationship matches our photometric detection
threshold of 26 mag~arcsec$^{-2}$, corresponding to a physical detection threshold 
of 3.2 $\times$ 10$^{-3}$ M$_{\odot}$pc$^{-2}$ or 0.12 M$_{\odot}$
per pixel.
There is, however, no obstacle to detecting a stellar component younger than few Myr 
and as massive as the older burst. The fact that such high, stellar mass surface densities 
are not derived for the younger burst
may hint at a constant star-formation rate over the last $\sim$10$^8$ yrs, so that the younger
burst has not had time yet to build-up all the stellar mass assembled by the older burst. 

From all the pixels enclosed in an ellipse of semi-major axis {\it a} = 8 kpc and semi-minor
axis {\it b} = 3.5 kpc (PA = 55$^o$) we derive a total mass of about 4.4 $\times$ 10$^7$ 
M$_{\odot}$ for the stars younger than 1 Gyr (from the fits that include the reddening
correction), of which $\simeq$ 3 $\times$ 10$^6$ M$_{\odot}$ were produced during the younger
burst and $\simeq$ 4.1 $\times$ 10$^7$ M$_{\odot}$ during the older burst. Switching off the 
reddening correction in our dating technique gives lower values: a total mass of about
3.9 $\times$ 10$^7$ M$_{\odot}$ split between $\simeq$ 2 $\times$ 10$^6$ M$_{\odot}$ and
$\simeq$ 3.7 $\times$ 10$^7$ M$_{\odot}$ for the younger and older burst, respectively. 

Although a SSP based on the assumption of istantaneous burst does not provide a true star-formation
rate, we can estimate one from the ratio of the total stellar mass produced
during a burst to the mean stellar age of the same burst. This exercise is based on the 
pixel-by-pixel maps of stellar age and mass surface density computed by omitting the reddening 
correction, so that we can compare the results with the SFR derived in the 
literature from the galaxy observed H$\alpha$ flux (i.e. not corrected for intrinsic reddening).
Therefore, for a mean stellar age of $\simeq$ 4 Myr and 150 Myr of the younger and older
burst, respectively, and the burst total masses as listed above, we  derive a SFR 
of about 0.5 M$_{\odot}$yr$^{-1}$ and 0.2 M$_{\odot}$yr$^{-1}$. Given that the typical 
uncertainty in our age and mass estimates is a factor of 2, the error on the SFR corresponds
to about a factor of 3, and within this uncertainty we may infer that the SFR has been nearly
constant over the last $\sim$10$^8$ yr in IC~2574. For comparison, the SFR derived from
the H$\alpha$ flux by Miller \& Hodge (1994) is $\simeq$ 0.1 M$_{\odot}$yr$^{-1}$.
This value suffers its own uncertainties 
concerning the depth of the observations as well as the escape fraction of ionizing photons
emitted by young and massive stars, which can be as high as 50$\%$ (cf. Weedman 1991, Pasquali
\& Castangia 2008). 

As already noted in Sect. 1 and 3, the Northeast complex of HII regions, also associated with
a HI supergiant shell, is the brightest site of star formation in IC~2574. In order to
investigate whether its stellar populations are different from the rest of the galaxy, we
analyze the distributions of pixel stellar ages and mass surface densities within the supergiant 
shell. These are plotted in Figure 9 as a function of distance
from the shell center. Here, each 2D bin is
0.4 dex $\times$ 28 pc in size. The shell is spatially rather uniform in age, except
perhaps for ages younger than 3 Myr (Log$_{10}$(AGE) = 6.5), which show up only at radii larger 
than 400 pc. It appears
that the most significant episode of star formation occurred about 5 Myr ago; a previous
burst may be identified in a more modest peak of the distribution at an age of $\sim$ 20 Myr.  
As discussed by Walter et al. (1998) and Walter \& Brinks (1999), the shell contains a star 
cluster and has 
9 star-forming regions distributed along its rim. Stewart \& Walter (2000) estimated an 
age of 11 Myr for the star cluster and 3 Myr for the star-forming regions. From the LBT data we
derive an age of about 5 Myr for the cluster and  1 - 3 Myr for the shell rim. These values
do not change significantly when the correction for intrinsic reddening is omitted or the
diffuse large-scale disk is not removed. The discrepancy in the cluster age between this
work and Stewart \& Walter could be related to different IMF and evolutionary tracks, 
and to a different treatment of reddening which can be quite patchy within the shell. However, 
given the factor of 2 uncertainties in our age estimates, 
these two results are consistent. The stellar mass 
surface density of the shell varies between 0.01 and
1 M$_{\odot}$pc$^{-2}$, and peaks at about 0.1 M$_{\odot}$pc$^{-2}$.

\subsection{Comparison with the HI column-density map}
Figure 10 shows the distribution of pixel stellar ages and mass surface densities as a 
function of the HI column density corrected for inclination. Here, each 2D bin is 0.4 dex 
$\times$ 0.2 dex in size. No correlation arises from
these plots except for the fact that the star-formation activity in IC~2574 seems to preferably
occur in regions with $\Sigma_{\rm HI} \simeq$ 6 $\times$ 10$^{20}$ cm$^{-2}$. This value is
comparable with the critical density derived by Skillman (1987, $\sim$ 10$^{21}$ cm$^{-2}$)
for a number of irregular galaxies and Kennicutt (1989, $\sim$ 5 $\times$ 10$^{20}$ cm$^{-2}$)
for a sample of disk galaxies. 
Column densities smaller than $\sim$10$^{20}$ cm$^{-2}$ are measured in 
correspondence with HI holes and may be not representative of the local HI gas 
prior to the formation of the holes. 
 
The lack of any correlation between star-formation and HI gas on local scales,
was also observed in seven spiral galaxies by Wong \& Blitz (2002), in NGC~6822 by
de Blok \& Walter (2006), in M~51a by Kennicutt et al. 
(2007) and for a large galaxy sample by Bigiel et al. (2008). In the case of M~51a, 
Kennicutt et al. showed that 
the local star formation density correlates with the local H$_2$ density
(and thus the local HI$+$H$_2$ density). In IC~2574, we expect that most the ISM
is atomic, and we might expect a relationship between HI and SFR based on
both formation of and feedback from young stars. From formation, the
correlation observed between H$_2$ and SFR in gas-rich spirals might suggest
that the gas reservoir, here probably HI, dictates the SFR. From feedback,
we would expect a correlation if HI forms via photodissociation of H$_2$
powered by the stellar UV radiation field (Shaya \& Federman 1987, Tilanus \& Allen
1991, Allen et al. 1997) or an anticorrelation if the
stellar radiation field ionizes HI into HII. None of these predictions
appear to be borne out by the results seen in Figure 10. We are then led
to conclude that there is no evidence in our data that HI either sets the
star formation rate or that star formation creates or ionizes a
significant fraction of the HI.

\begin{figure}
\epsscale{0.80}
\plotone{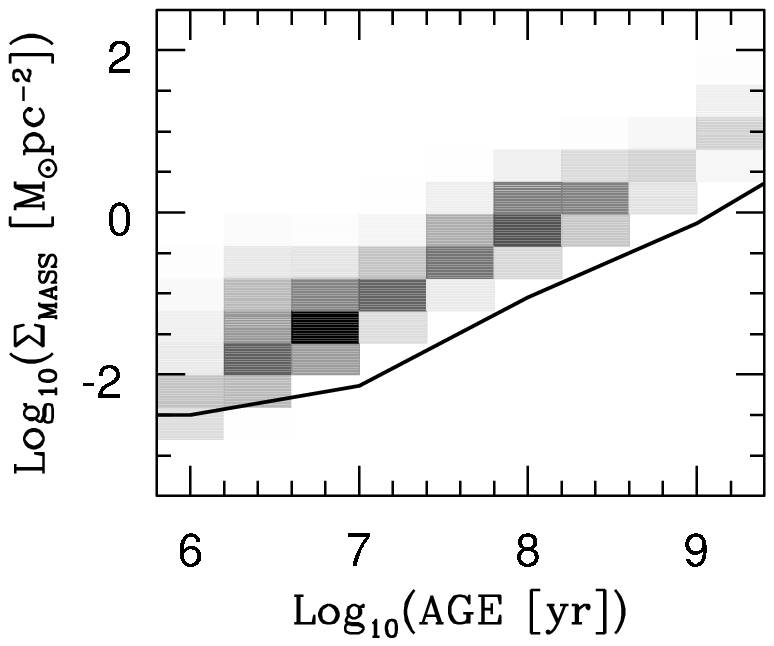}
\caption{The distribution of  stellar mass surface density (corrected for inclination) as a
function of stellar age (obtained after correction for reddening). The grey scale is based
on the number of pixels within each 2D bin,
0.4 dex $\times$ 0.4 kpc in size, and it grows darker as this number increases. The
solid black line corresponds to a detection threshold of 3.2 $\times$ 10$^{-3}$
M${_\odot}$pc$^{-2}$ or 0.12 M${_\odot}$ per pixel.}
\end{figure}

\section{Stellar properties of the HI holes}
The lack of correlation between HI and star formation is especially
striking in IC~2574, where the HI distribution is marked by a wealth of
holes. A natural explanation for these structures is that they
are carved out by feedback from star formation, but the lack of a
correlation between atomic gas and the SFR calls such a relationship into
question. In this section we ask whether young stars have been able to
produce the observed HI holes and to sustain their expansion; we also ask
whether such holes are a natural consequence of the star formation we
observe and investigate the implications of the lack of HI-SFR correlation
for stellar feedback.

\begin{figure*}
\epsscale{0.80}
\plotone{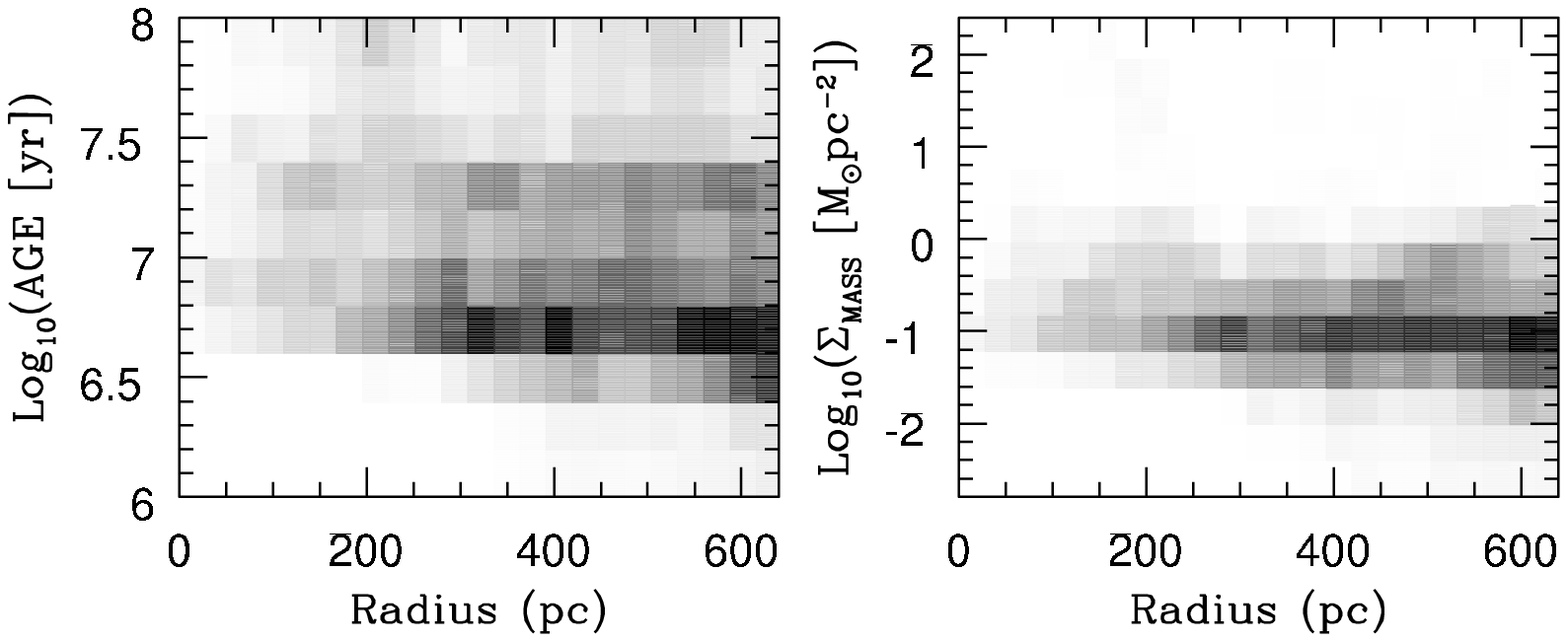}
\caption{The distributions of pixel stellar age and mass surface density as a function 
of pixel distance from the center of the supergiant shell in the Northeast complex of HII
regions. Stellar mass surface densities are corrected for inclination. Stellar ages and
mass surface densities are obtained from the fits which apply the correction for reddening.}
\end{figure*}

\begin{figure*}
\epsscale{0.80}
\plotone{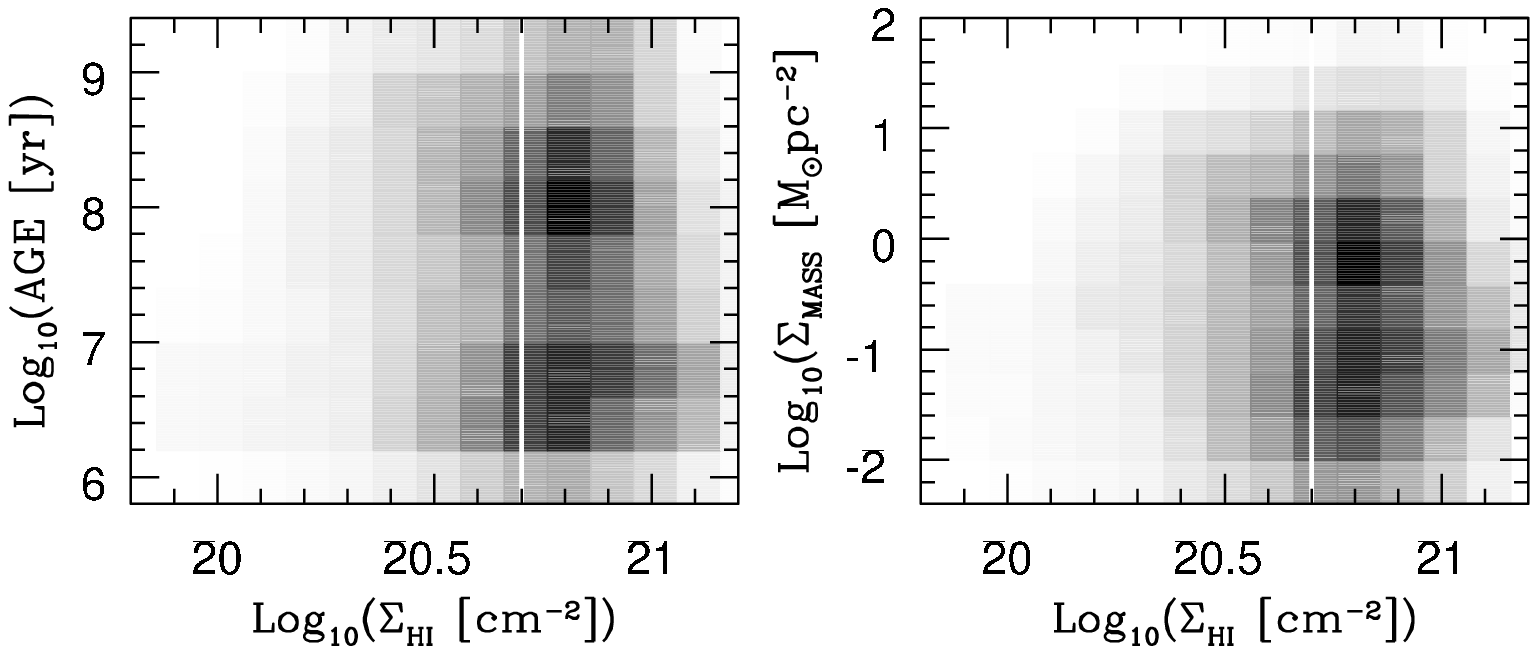}
\caption{The pixel stellar ages and mass surface densities as a function of HI column density corrected
for inclination. Each 2D bin is 0.4 dex $\times$ 0.1 dex in size, and is color coded on the basis
of the number of pixels falling within it. The grey shade becomes darker with increasing number of
pixels. The thick white, vertical line indicates the characteristic critical density obtained by
Kennicutt (1989), $\sim$ 5 $\times$ 10$^{20}$ cm$^{-2}$.}
\end{figure*}

\subsection{Are the HI holes powered by stars?}

Walter \& Brinks (1999) identified 48 holes and measured their diameter, expansion 
velocity, dynamical age, volume/column density and their HI mass. These holes are long-lived
features; since IC~2574 largely rotates as a solid body, the shear computed from its 
smoothed rotation curve turns out to disrupt the HI holes over timescales larger than 0.5 Gyr. 
Out of the initial 48 holes, only 15 
overlay with the residual images of IC~2574 obtained after subtracting the smooth disk
component (see Sect. 4), have a measured velocity expansion, and more than 80$\%$ of 
their area filled with pixels carrying information on stellar age and mass density. 
Similarly to the supergiant shell whose star cluster is off-centered (Walter \& Brinks,
Stewart \& Walter 2000), we do not detect any central star cluster in these 15
HI holes, but rather a spatial mixture of populations formed during the younger and
older bursts. The spatial distribution of the younger population within each HI hole
can be clumpy, with the clumps often located at the periphery of the HI hole.
Figure 11 shows the radial distribution of pixels with an associated stellar age
younger than 16 Myr (corresponding to the younger burst) obtained for each HI hole.
The three panels of Figure 11 distinguish the holes completely filled by young 
stellar populations (left-hand side panel) from those where young stellar ages
are found in the periphery (middle panel) and from the holes where 
young stellar populations are preferentially central (right-hand side panel). 
We express the distance of each pixel from the hole center in 
units of the respective hole radius and compute the number of pixels younger than 16 Myr per
distance bin as well as the total number of pixels younger than 16 Myr within
the hole. We use this latter quantity to normalize the former, and derive in this
way the fraction of ``young'' pixels as a function of distance within a HI hole. 
Apart from the three holes in the left-hand side plot that are completely dominated by a
young stellar component, in most cases (including the supergiant shell)
this young stellar population lives in the outer skirts of a hole, at a distance
larger than 0.4 times the hole radius (central panel). Only in two holes does the
young component look more centrally concentrated (right-hand side plot). 

For these 15 holes we also estimate: {\it i)} the fraction of pixels associated 
with a stellar age younger than 16 Myr ($f_{\rm YOUNG}$), defined as the ratio of pixels
associated with stellar ages younger than 16 Myr to the total number of pixels within a hole;  
{\it ii)} the mean age of the stellar populations produced during the younger burst (with
an age $<$ 16 Myr), {\it iii)} the total stellar energy released by the stellar winds and 
supernova explosions of the young population
($E_{\rm YOUNG}$), and finally {\it iv)} the kinetic energy of the HI holes in terms of a fraction
of the total energy provided by their young stellar populations. We compute the stellar
total energy, $E_{\rm YOUNG}$, from the value derived by STARBURST99 
at the mean age of the young
component within a hole and properly scaled by its total mass. The kinetic energy
of a hole is determined from its column density, corrected for inclination ($\Sigma_{\rm HI}$
estimated by Walter \& Brinks) and its expansion velocity ($V_{\rm exp}$): 
$E_{\rm HI}$ = 0.5$\Sigma_{\rm HI} \times Area \times V_{\rm exp}^2$, 
where $Area$ is the area of the hole
whose diameter is listed in Walter \& Brinks. We then calculate the ratio of the hole
kinetic energy to the stellar total energy $E_{\rm HI}/E_{\rm YOUNG}$; this ratio 
is a measure of the efficiency at which the total stellar energy is transformed into kinetic 
energy of the HI gas.

\begin{figure*}
\epsscale{1.00}
\plotone{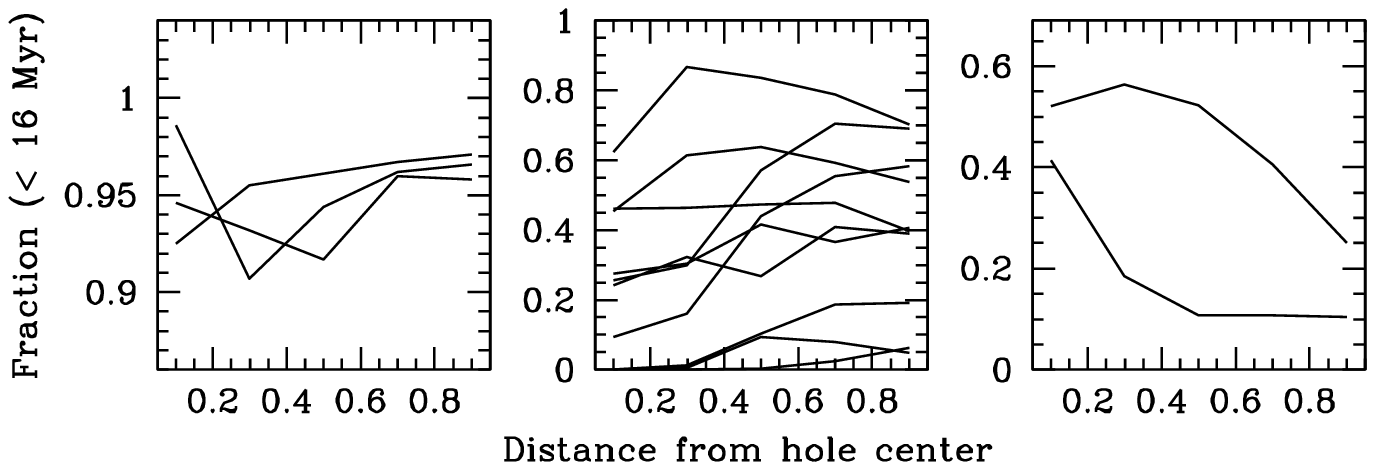}
\caption{The fraction of pixels associated with a stellar age younger than 16 Myr as a function
of distance from the hole center. The distance is units of the respective
hole radius. The number of ``young''
pixels per bin of distance is normalized by the total number of pixels younger than 16 Myr.
{\it Left-hand side panel:} holes filled with young stellar populations (3 of 15).
{\it Middle panel:} holes with young stellar populations along the outskirts ( 10 of 15).
{\it Right-hand side panel:} holes with centrally concentrated young stellar populations
(2 of 15).}
\end{figure*}

All parameters mentioned above are plotted in Figure 12. Panel {\it a} presents $f_{\rm YOUNG}$ 
as a function of the hole diameter in parsec. No correlation is seen between these two parameters.
The mean age of the
young stellar component is plotted as a function of the hole dynamical age in panel 
{\it b}. Here there is a hint of a relation in which both the young component and the hole
age together, although the stars generally appear to be younger than their associated 
hole by a factor of 5 on average.  We note here that
the dynamical age of the holes is an upper limit to the true hole age, since Walter \&
Brinks derived the dynamical age under the assumption of a constant expansion velocity,
while these holes have certainly expanded faster in the past. 
Because of the uncertainty of this assumption and the errors in our dating technique,
it is difficult to establish how much older (if at all) the HI holes are compared to  
their young stellar component.
As shown in panel {\it c}, the total energy released by the young stellar component is 
typically an order of magnitude larger than the kinetic energy of the HI holes and 
lies above the one-to-one 
correspondence traced by the grey line. Indeed, the ratio of the hole kinetic energy to 
the total stellar energy is on average about 10$\%$, thus indicating that the
hole expansion can be sustained with only 10$\%$ of the available total stellar energy 
(cf. panel {\it d}).
This overall picture does not qualitatively change when we use the stellar populations
obtained by switching off the reddening correction in our dating technique. In this
case, the ratio of the hole kinetic energy to the total stellar energy increases 
from 10$\%$ to about 20$\%$, since the output stellar mass surface density is lower when reddening 
is neglected. An $E_{\rm HI}/E_{\rm YOUNG}$ ratio of $\sim$10 - 20$\%$ is in agreement 
with the value (10$\%$ - 20$\%$) independently obtained
by Cole et al. (1994) from modeling the galaxy luminosity function, while Bradamante
et al. (1998) determined a value of 3$\%$ for supernovae II and 100$\%$ for supernovae Ia from
the modeling of the chemical evolution of blue compact galaxies. 

\begin{figure*}
\epsscale{0.70}
\plotone{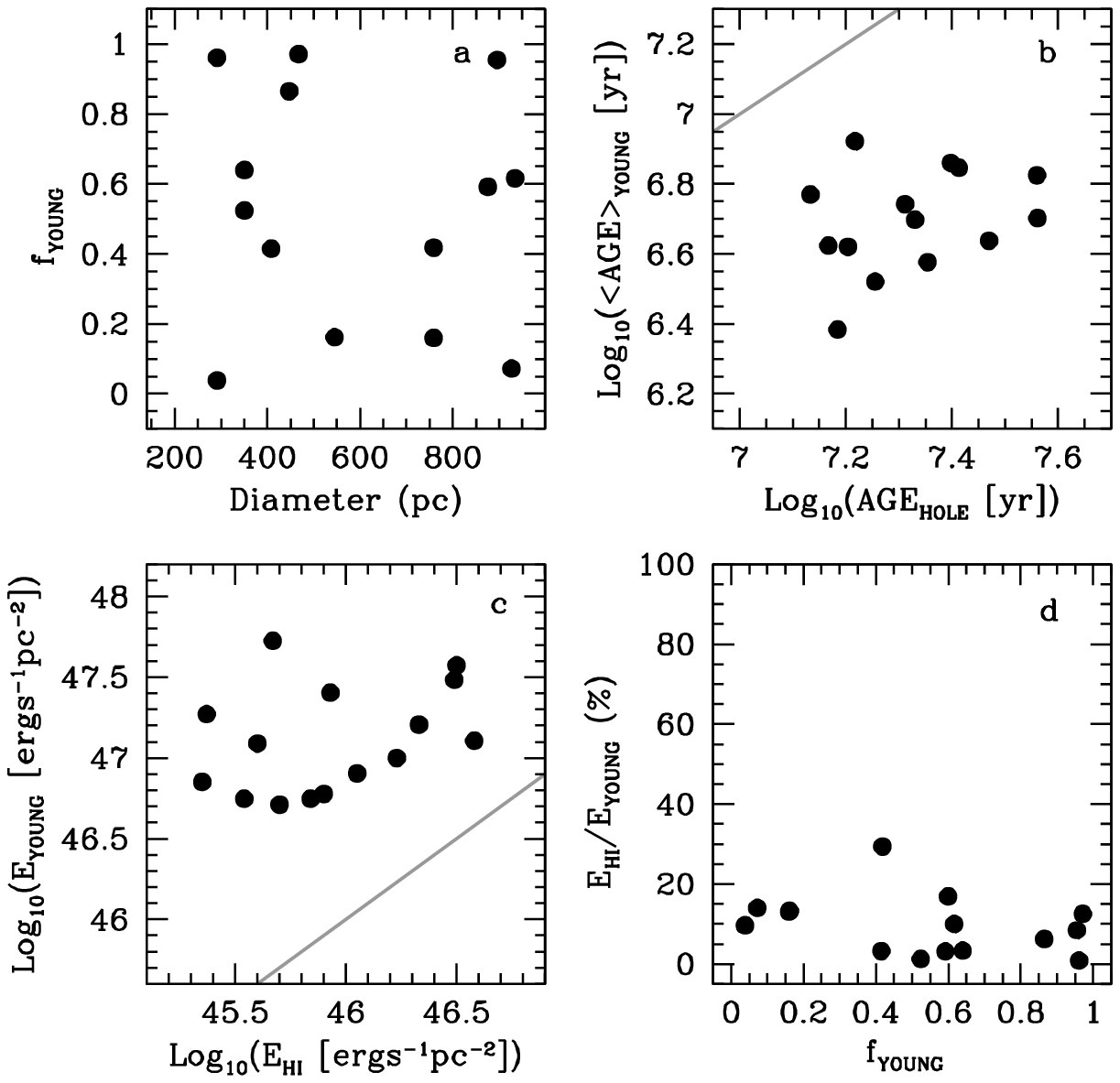}
\caption{The total energy budget of the younger burst ($<$ 16 Myr) detected
within 15 HI holes. {\it Panel a:} the ratio of pixels associated with stellar ages younger
than 16 Myr to the total number of pixels within a hole is plotted as a function of the hole
diameter, in parsec. {\it Panel b:} the mean age of the young stellar component ($<$ 16 Myr)
enclosed in a HI hole is shown as a function of the hole dynamical age, simply computed from its
diameter and expansion velocity. The holes stellar and dynamical ages become equal along
the grey line. {\it Panel c:} the total energy released by the stellar winds
and supernova explosions of the young component is traced as a function of the hole
kinetic energy, derived from the hole expansion velocity. Both quantities are normalized
by the hole area. The grey line indicates
when the total stellar energy equals the hole kinetic energy. {\it Panel d:} based on
panel {\it c}, the ratio of
the hole kinetic energy to the total energy of its young component is plotted versus
the fraction of pixels associated with stellar ages $<$ 16 Myr.  }
\end{figure*}

\subsection{Why doesn't all star formation create holes?}
If a conclusion from Figure 12 is that the stellar energy budget is in principle large enough to
support the holes expansion, we may then ask why many star-forming regions in IC~2574
are not located in HI holes. We build a sample of 105 control apertures with the
diameter (7 arcsec or 136 pc at the assumed distance of IC~2574) of the beam
of the HI observations, placed randomly across the galaxy to sample
different stellar populations and different values of the HI column density away from
the observed HI holes. 
For each aperture, we measure $f_{\rm YOUNG}$, $E_{\rm YOUNG}$, $\Sigma_{\rm HI}$,
the average HI velocity dispersion, $\sigma_{\rm HI}$ (from the HI second-moment map of
Walter \& Brinks), $E_{\rm HI}$ defined as 1.5$\Sigma_{\rm HI} \times Area \times \sigma_{HI}^2$
(where $Area$ is the area of the apertures), and the energy ratio $E_{\rm HI}/E_{\rm YOUNG}$. 
All these properties are plotted in grey in Figure 13 for only those 37 apertures with 
$f_{\rm YOUNG} >$ 70$\%$, so that their stellar 
populations mostly formed during the younger burst. In black are the observed
15 HI holes, whose $\sigma_{\rm HI}$ we measure in an annulus about 100 pc wide around
each hole and whose $\Sigma_{\rm HI}$ was estimated by Walter \& Brinks.  
The assumption here is that the HI properties outside a HI hole may be representative of the
local gas before the hole formation. Panel {\it a} of Figure 13 shows the total stellar
energy $E_{\rm YOUNG}$ as a function of the HI kinetic energy, 
both parameters normalized by the area of the apertures (grey circles)
and HI holes (black circles). The grey line indicates the one-to-one correspondence between these
two energy terms. For the majority of apertures as well as for all the HI holes, the total stellar 
energy is up to an order of magnitude larger than the HI kinetic energy. Indeed, the
$E_{\rm HI}/E_{\rm YOUNG}$ ratio is lower than unity for about 70$\%$ of the control apertures,
and the average $E_{\rm HI}/E_{\rm YOUNG}$ is $\simeq$ 35$\%$, indicating that about one third
of the total stellar energy can account for the observed HI kinetic energy (cf. panel {\it b}).
Yet, no hole in the atomic H gas is detected at the position of these apertures. We notice that the
apertures whose $E_{\rm HI}/E_{\rm YOUNG}$ is larger than 100$\%$ are systematically associated with
$\Sigma_{\rm HI} \geq$ 8 $\times$ 10$^{20}$ cm$^{-2}$, i.e. high HI columns. The energy ratio 
for the HI holes is
well below 100$\%$, with an average value of about 10$\%$; all HI holes have $\Sigma_{\rm HI} <$
8 $\times$ 10$^{20}$ cm$^{-2}$. In panel {\it c} of Figure 13 the total stellar energy is
compared with the average velocity dispersion for both the control apertures (grey circles)
and the observed 15 HI holes (black circles). No clear correlation exists between these
two parameters, although one could attempt to see $E_{\rm YOUNG}$ increasing with $\sigma_{\rm HI}$.    
This trend is seen also across a large sample of galaxies by Tamburro (2008) and could be 
interpreted as due to star formation inducing turbulence in the HI gas (cf. also Dib et
al. 2006).  
The total stellar energy is plotted as a function of the HI column density, $\Sigma_{\rm HI}$,
in panel {\it d}; once again, no tight correlation can be seen in this plot, except 
perhaps for a weak trend whereby $E_{\rm YOUNG}$ decreases at larger $\Sigma_{\rm HI}$. 

As pointed out by Walter \& Brinks, the HI velocity contours of IC~2574 are characterized 
by distortions due to non-circular motions. Non-circularity is believed to arise from
star-formation processes or from the presence of a bar, a spiral density wave or
non-circular halo potentials. Oh et al. (2008) have recently isolated the non-circular
motions in IC~2574 and showed that the typical velocity of non-circular motions is few
km~s$^{-1}$, with peaks of $\sim$ 20 km~s$^{-1}$. We use the map of the kinetic
energy associated with non-circular motions constructed by Oh et al. 
(as $E_{\rm HI}^{\rm ncir}$ = 0.5 $\times$ M$_{ncir} \times$ V$_{ncir}^2$, where
M$_{ncir}$ is the HI mass associated with non-circular motion, see Oh et al. for
further details)  and compute the
total kinetic energy in non-circular motions in the control 
apertures with $f_{\rm YOUNG} >$ 70$\%$ as well as in the HI holes. 
We compare this energy with the total stellar energy in the left-hand side
plot of Figure 14, where the black and grey circles represent the HI holes and the control 
apertures, respectively, and the grey solid line traces the one-to-one correspondence
between the kinetic non-circular and stellar energies. The large majority of control
apertures and holes are placed above this line indicating that the energy released
by their young stellar population is larger than the kinetic energy of the in-situ 
non-circular motions.  
The distribution of the control apertures (in grey) and hole
(in black) in the energy ratio $E_{\rm HI}^{\rm ncir}/E_{\rm YOUNG}$ is plotted in
the right-hand side panel of Figure 14. 
Values of Log$_{10}(E_{\rm HI}^{\rm ncir}/E_{\rm YOUNG}) <$ 0 are obtained for those
control apertures and holes where the total stellar energy is in excess of
$E_{\rm HI}^{\rm ncir}$; they define an average $<E_{\rm HI}^{\rm ncir}/E_{\rm YOUNG}>$ = 
21$\%$ (control apertures) and 43$\%$ (holes).

\begin{figure*}
\epsscale{0.70}
\plotone{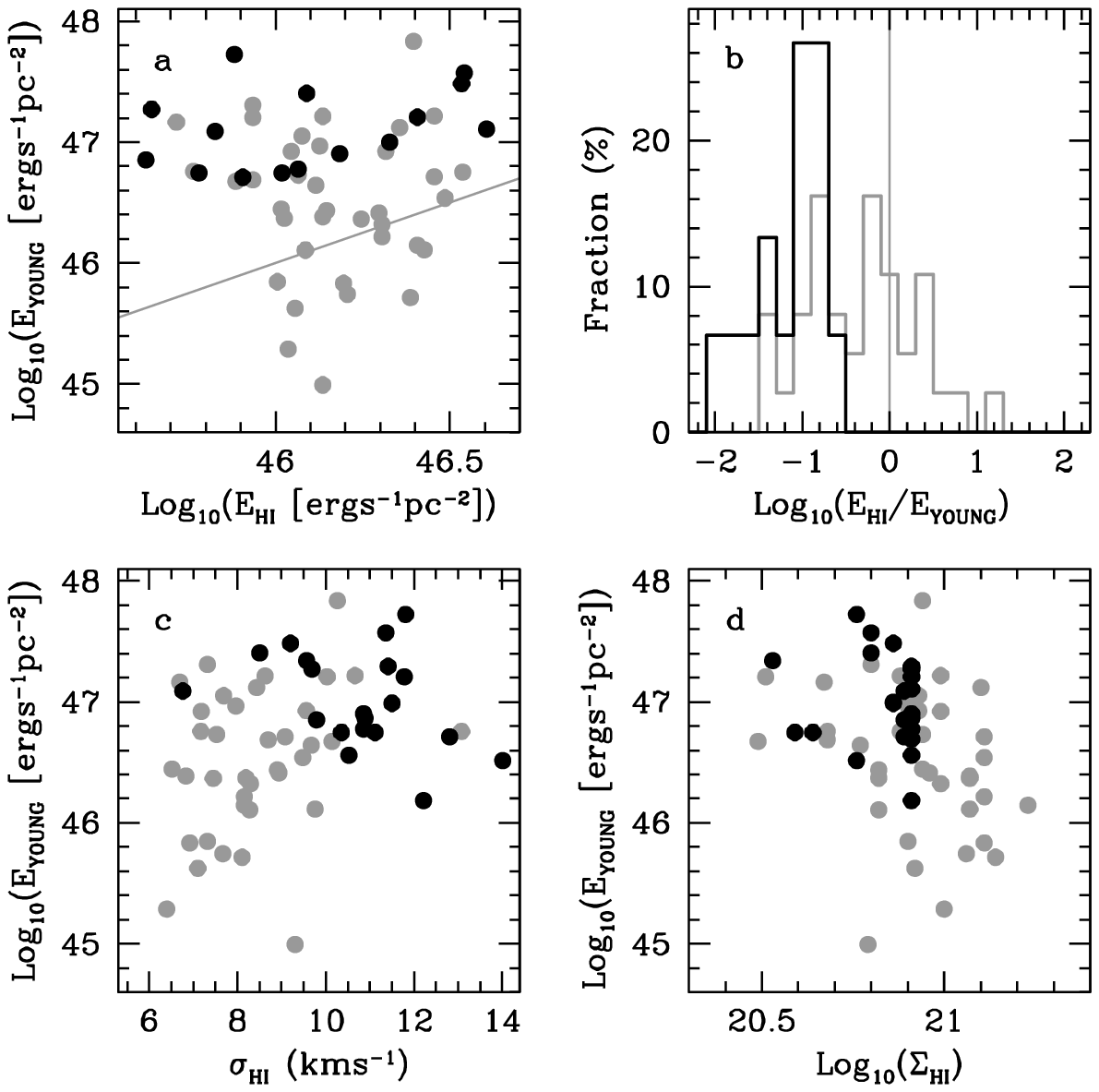}
\caption{{\it Panel a:} the comparison of the total energy released by the stellar populations
younger than 16 Myr with the HI kinetic energy based on the velocity dispersion of the atomic
H gas. Black and grey circles represent 15 HI holes and the control apertures, respectively.
The latter are characterized by a fraction of pixels associated with stellar ages $\leq$ 16
Myr larger than 70$\%$. The grey line indicates the one-to-one correspondence between $E_{\rm YOUNG}$
and $E_{\rm HI}$. Both energy terms are normalized by the area of the holes and the control
apertures. {\it Panel b:} the distribution of the energy ratio $E_{\rm HI}/E_{\rm YOUNG}$, in black
for the HI holes and in grey for the control apertures. The vertical grey line highlights
the $E_{\rm HI}/E_{\rm YOUNG}$ value of 100$\%$. {\it Panel c:} $E_{\rm YOUNG}$ as a function of the HI
velocity dispersion, $\sigma_{\rm HI}$. {\it Panel d:} $E_{\rm YOUNG}$ as a function of the HI
column density, $\Sigma_{\rm HI}$.}
\end{figure*}
 
\subsection{Photoionization of HI}
There is a final aspect to consider in our analysis of the HI holes formation, i.e. the
ionization of the HI gas due to stellar radiation. So far, we have considered the total
energy produced by the stellar winds and supernova explosions of the stellar population 
younger than 16 Myr. 
Figures 12 and 13 do not show any firm evidence for a stellar origin of the
holes, although the HI kinetic energy of the holes and the control apertures is typically
about 10 - 30$\%$ of the total stellar energy. We now ask whether the HI holes are produced 
by the stellar radiation field via ionization of the atomic H gas. We
estimate the number of ionizing photons, $Q(H)$, in each hole and control aperture from 
the value tabulated by STARBURST99 at the age of the young component in the hole/aperture
scaled by its total mass. We then use $Q(H)$ to compute the radius, $R_S$, of the 
Str\"omgren sphere associated with each HI hole and control aperture. The Str\"omgren sphere defines
the volume within which all the HI gas is ionized, and its radius $R_S$ is
expressed by the equation $Q(H)$ = (4$\pi R_S^3 n_{\rm HI}^2 \alpha_B )/$3, where $n_{\rm HI}$ is 
the HI volume density and
$\alpha_B$ a scaling parameter depending on the gas temperature. From Osterbrok (1989) we
assume $\alpha_B$ = 4.54$\times$10$^{-13}$ for a gas temperature of 5000 K. We 
derive the HI volume density $n_{\rm HI}$ from the relation: 
$\Sigma_{\rm HI}$ = $\sqrt{2 \pi} h n_{\rm HI}$
where $h$, the scale height of the HI disk, is 440 pc as obtained by Walter \& Brinks (1999)
for a distance of 4.02 Mpc. 
\par\noindent
If all the stellar photons produced within the control apertures
ionized the surrounding HI gas, the radius of the Str\"omgren sphere of
the control apertures with $f_{\rm YOUNG} >$ 70$\%$  would be on average
4 times bigger than the aperture radius, and the apertures would lie in holes that are well
observable. Given that the control apertures are not associated with HI holes by definition,
only a much lower fraction of $Q(H)$ can ionize the HI gas.
This limits the number of ionizing photons to less than 20$\%$ of $Q(H)$; under this
assumption, the ratio of $R_S$ to the aperture radius decreases to an average value of 2, 
consistent with an uncertainty on $Q(H)$ of a factor of 3 which is due to the errors on the 
stellar ages and masses. This constraint would make any HI hole formed via ionization less 
likely detectable. For the HI holes with $f_{\rm YOUNG} >$ 70$\%$, the UV radiation from 
the young stellar populations would produce a Stromgren sphere 4 times larger than the 
hole radius if 100$\%$ of the ionizing photons are available.  Making the observed hole 
consistent with the Stromgren radius requires than only $\sim 20\%$ of the available photons.
Therefore, such a limit on $Q(H)$ may be sufficient to 
explain both the
existence of those HI holes whose stellar populations are predominatly young ($<$ 16 Myr),
and the lack of observable HI holes around the control apertures. 

\begin{figure*}
\epsscale{0.80}
\plotone{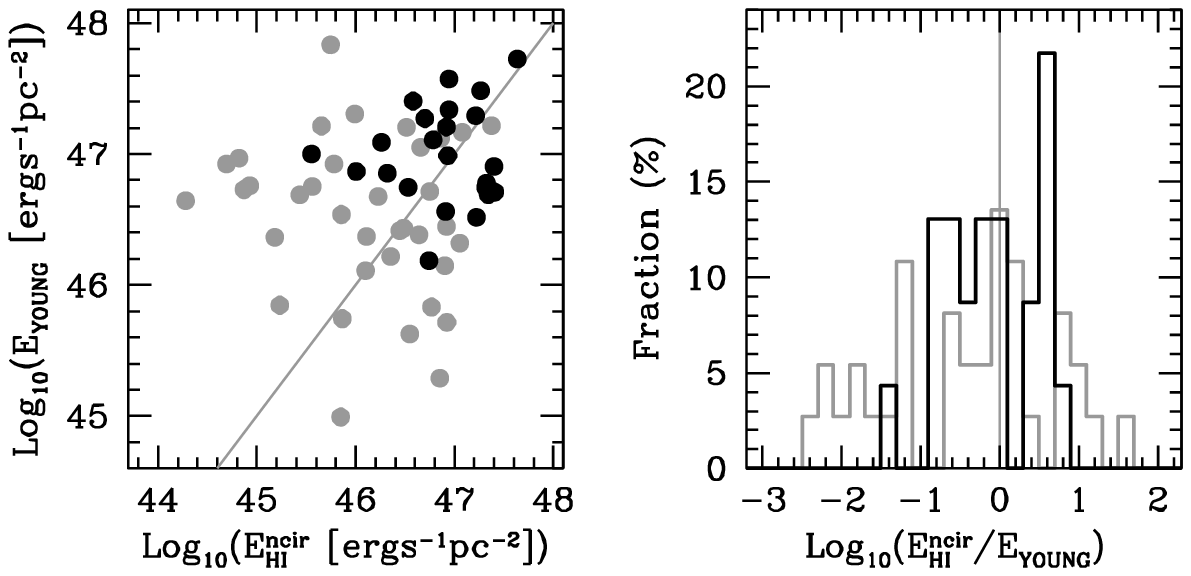}
\caption{{\it Left-hand side panel}: the comparison between the total energy released by the
stellar populations younger than 16 Myr, $E_{\rm YOUNG}$, and the HI kinetic energy associated
with non-circular
motions, $E_{\rm HI}^{\rm ncir}$. Black and grey circles represent the HI holes and the control
apertures, respectively. {\it Right-hand side panel}: the distribution of the control apertures
(in grey) and HI holes (in black) in the energy ratio $E_{\rm HI}^{\rm ncir}/E_{\rm YOUNG}$.
The grey vertical line indicates Log$_{10}(E_{\rm HI}^{\rm ncir}/E_{\rm YOUNG})$ = 0, i.e.
$E_{\rm HI}^{\rm ncir}/E_{\rm YOUNG}$ = 100$\%$.}
\end{figure*}

\section{Discussion and conclusions}
Deep, LBT imaging in the $U$, $B$ and $V$ bands has allowed us to spatially map the
star formation history of IC~2574 over the last $\sim$10$^8$ yrs. We do this by analyzing
the pixel-by-pixel colors of all areas in IC~2574 with significant detections in all
bands. Overall, the angle averaged surface-brightness profiles 
indicate a predominance of blue colors [($U - B$) $<$ -0.3 mag and ($B - V$) $<$ 0.4 mag] 
at galactocentric radii larger that 4 kpc, mostly due to the presence of the large
North-East complex of HII regions (associated with a HI supergiant shell) and 
the South-West extended tail of star formation.
At a limiting brightness of 26 mag~arcsec$^{-2}$, the major axes of the HI and $U$-band 
light distributions are nearly the same, while the minor axis is smaller in the optical. 
This could be due to differences in the scale height of the $U$-band light and HI 
distributions.

Using the dating technique of Pasquali et al. (2003), we are  able to construct 
pixel-by-pixel maps of stellar age and mass surface density from the comparison of the observed 
colors with simple stellar population models synthesized with STARBURST99. After fitting
and subtracting a smooth disk component (whose colors point to a stellar age $>$ 200 Myr), 
our best-fitting STARBURST99 models indicate the occurrence of two major bursts, one about 
100 Myr ago and the other during the last 10 Myr. The older burst appears to be spatially
``confined'' within the inner 4 kpc (in galactocentric radius), while the younger episode
of star formation occupies the 4 $< R <$ 8 kpc region. The two episodes differ also in
the stellar mass surface density of stars formed: the typical mass density associated with the older 
burst is about 1 M$_{\odot}$pc$^{-2}$, while it is $\simeq$0.04 M$_{\odot}$pc$^{-2}$ 
for the younger burst. However, 
these values are luminosity-weighted and therefore biased. Given that the $U$-band flux 
emitted by a single stellar population decreases for increasing age and the flux itself is 
proportional to the stars total mass, the detection of stars above a fixed threshold in
the $U$-band (used in this study as a prior) requires stellar populations progressively
more massive as they grow older. Quantitatively speaking, a limiting surface brightness of
26 mag~arcsec$^{-2}$ allows us to detect stellar populations older than 10$^8$ yrs with
a mass surface density larger than 0.1 M$_{\odot}$pc$^{-2}$, while those populations few Myr
young are detected down to a mass surface density of 0.003 M$_{\odot}$pc$^{-2}$ (a factor of 
$\sim$ 30 lower than at 100 Myr). The fact that we do not detect young stellar populations
as massive as 1 M$_{\odot}$pc$^{-2}$ is not a selection effect of the data; it possibly
hints at a constant star formation rate, at least over the last $\sim$10$^8$ yrs, 
so that the younger
burst has not had time yet to produce the same amount of stellar mass as was assembled during the
older burst.  

We compare both the stellar ages and the mass surface densities of the recently
formed populations  with the HI column density all across IC~2574 and do not find any
significant correlation between the star formation activity and
the HI reservoir in this galaxy. A similar result was obtained by Wong \& Blitz (2002),
de Blok \& Walter (2006), Kennicutt et al. (2007) and Bigiel et al. (2008), using  
different tracers of star formation. The lack of any correlation between star formation 
and HI is in contrast with the expectation of a relationship between HI and SFR based on
both formation of and feedback from young stars. In terms of star formation, the
correlation observed between H$_2$ and SFR in gas-rich spirals might indicate that
the HI gas reservoir controls the SFR. In terms of stellar feedback, instead, we would
expect to see a correlation if HI forms via photodissociation of H$_2$ powered by the stellar 
UV radiation field (Shaya \& Federman 1987, Tilanus \& Allen 1991, Allen et al. 1997) or
an anticorrelation if the stellar radiation field ionizes HI into HII. 
We thus conclude that the star formation in IC~2574 is not
regulated by the available HI gas reservoir, nor efficient at creating HI
via photodissociation, nor actively expeling or ionizing a significant
fraction of the HI.  On a global scale,  
star formation in IC~2574 appears to preferentially take place in regions where the HI column 
density is about 6 $\times$ 10$^{20}$ cm$^{-2}$. This value is comparable with the global critical 
density derived by Skillman (1987, $\sim$10$^{21}$ cm$^{-2}$) for a sample of irregular galaxies, by
Kennicutt (1989, about 5 $\times$ 10$^{20}$ cm$^{-2}$) for 33 disk galaxies, and by de Blok \& Walter
for NGC~6822 ((5.6 $\times$ 10$^{20}$ cm$^{-2}$), and with
the theoretical values computed by Schaye (2004). 

The lack of correlation between the star formation activity and the HI density in IC~2574 is 
rather puzzling with respect to the distinct HI holes, 
as one would expect to systematically detect a 
central cluster of young stars driving the hole expansion through their stellar winds and supernova 
explosions. With the possible exception of the supergiant shell (cf. Stewart \& Walter 2000), we do not 
observe any concentration of young stellar populations at the center of the HI holes, but 
rather a mixture of 
populations formed during the younger and older bursts, with the younger population often in clumps
located at the periphery of the HI hole. We
determine the fraction of pixels with a stellar age younger than 16 Myr ($f_{\rm YOUNG}$) for 15
of the 48 holes identified by Walter \& Brinks (1999) (i.e. those that overlap with the area of 
the galaxy obtained after the removal of its smooth disk component, see Sect. 4) . 
Here, the age of 16 Myr is assumed to separate the younger from the older star formation 
episode identified by us.  
We also compute the mean age ($<$AGE$>_{\rm YOUNG}$)
and the total energy (E$_{\rm YOUNG}$, due to stellar winds and supernova explosions) of 
the stars formed during the younger burst in each hole and
compare them with the dynamical age and kinetic energy of the hole (based on its expansion
velocity). The relevant results of such a comparison are that:
\par\noindent
{\it i)} $f_{\rm YOUNG}$ (the ratio of the number of pixels associated with stellar ages younger
than 16 Myr to the total number of pixels within a hole) does not correlate with the hole properties.
\par\noindent
{\it ii)} The dynamical ages of the holes are generally higher than the younger burst by a 
factor of 5. The holes dynamical age is though an upper limit as it was computed by
Walter \& Brinks assuming a constant expansion rate. In reality, the holes expansion is
expected to have been faster during its earlier phases.
\par\noindent
{\it iii)} The kinetic energy of the holes is on average 10$\%$ of the total energy 
released by the younger burst. This energy ratio (i.e. the efficiency of stellar feedback)
is in agreement with the value (10$\%$ - 20$\%$) obtained
by Cole et al. (1994) from fitting the galaxy luminosity functions and colors, while Bradamante
et al. (1998) determined a value of 3$\%$ for supernovae II and 100$\%$ for supernovae Ia from
the modeling of the chemical evolution of blue compact galaxies.
\par
Even at the remarkable photometric depth of the LBT data, we do not find a 
clear one-to-one association between the observed HI holes and the most recent bursts 
of star formation. 
However, the stars formed during the younger burst do in principle release enough energy to 
power the expansion of the HI holes. In order to investigate better this apparent 
discrepancy, we map IC~2574 in optical and in HI with a set of control apertures equivalent 
in size to the beam of the radio observations (136 pc). No observed HI hole is included among the 
control apertures. For those apertures with $f_{\rm YOUNG} >$ 70$\%$ (just to 
focus on the interplay
between the younger stellar component and the HI gas), we measure their average HI velocity
dispersion $\sigma_{\rm HI}$ (from the second-moment map of Walter \& Brinks), their 
average HI column density
$\Sigma_{\rm HI}$, the total stellar energy of their stellar populations younger than 16 Myr, 
$E_{\rm YOUNG}$, and the HI kinetic energy $E_{\rm HI}$, defined as $E_{\rm HI}$ = 
1.5$\Sigma_{\rm HI} \times Area \times \sigma_{\rm HI}^2$
(where $Area$ is the area of the apetures). For two thirds of the control apertures the ratio
$E_{\rm HI}/E_{\rm YOUNG}$ is lower than 100$\%$ and has an average value of 35$\%$; 
all the remaining 
apertures have $E_{\rm HI}/E_{\rm YOUNG} >$ 100$\%$ and are systematically associated with 
$\Sigma_{\rm HI} \geq$ 8 $\times$ 10$^{20}$ cm$^{-2}$. Once again, the stellar populations 
formed during the younger burst 
could balance the kinetic energy stored in the HI gas, transform it into an expansion motion and
thus produce a detectable HI hole for most of the control apertures. Yet, 
these regions failed to develop a HI hole. These same stellar populations could also
sustain the HI non-circular motions, since their average ratio $E_{\rm HI}^{\rm ncir}/E_{\rm YOUNG}$ 
is 21$\%$. 

Holes in the HI distribution may also be formed by ionization due to the UV emission of the
young stars. In the case of those control apertures with 
$f_{\rm YOUNG} >$ 70$\%$, the Str\"omgren sphere produced by their stellar populations younger 
than 16 Myr
is 4 times bigger in radius than the apertures and, therefore, would produce a detectable HI
hole. Since this is not observed, the fraction of ionizing photons that effectively ionize the HI 
gas has to be rather small; a fraction $\leq$ 20$\%$ of the total number of ionizing photons emitted by
the young stellar populations reduces the radius of the Str\"omgren sphere to $\leq$ 2 times the 
aperture radius, making any HI hole formed via ionization less likely detectable. This result is in
agreement with the lack of correlation between star formation activity and HI density seen in Sect. 4.3,
indicating that star formation does not appear to significantly affect the properties of the HI gas in
IC~2574. From the analysis of the stellar content of the HI holes and the control apertures 
it is hard to establish whether stellar/supernova feedback is at work in IC~2574 on local scales of 
few 100 pc; this same result has been found for other dwarf galaxies (LMC and SMC, 
Kim et al. 1999, Hatzidimitriou et al. 2005; M33, van der Hulst 1996; Holmberg II, Rhode et al. 1999).  
The ultimate explanation of these findings remains elusive; it may rest either in
an overestimate of the stellar mass loss and stellar winds especially at low metallicity, or in our
understanding of how stellar energy interacts with
the interstellar medium. Another explanation may be found in a stochastic sampling of the IMF in 
low-mass stellar systems. Stellar-populations synthesis codes usually assume that the IMF
is a densely sampled probabilistic distribution function, while the IMF is in reality poorly
and discretely sampled in low-mass stellar systems. This assumption produces fluctuactions
in the predicted properties of these systems which may mislead comparisons with the observations
(Cervi\~no \& Valls-Gabaud 2003), by overestimating, for example, the stellar energy realesed 
to the ISM and thus underestimating the efficiency of stellar feedback.
As shown by Cervi\~no \& Molla (2002), Cervi\~no \& Luridiana
(2006) and Carigi \& Hernandez (2008), stochastic effects on the IMF sampling become significant 
in low metallicity systems such as dwarf galaxies.

\acknowledgments
We would like to thank S.-H. Oh for making available the map of the kinetic energy associated
with non-circular motions in IC~2574, and P. Smith for assisting with the observations.
Based on data acquired using the Large Binocular Telescope (LBT).
This research made use of tools provided by Astrometry.net
The LBT is an international collaboration among institutions in the United States,
Italy and Germany. LBT Corporation partners are: The University of Arizona on behalf
of the Arizona university system; Istituto Nazionale di Astrofisica, Italy; LBT
Beteiligungsgesellschaft, Germany, representing the Max-Planck Society, the
Astrophysical Institute Potsdam, and Heidelberg University; The Ohio State University,
and The Research Corporation, on behalf of The University of Notre Dame, University of
Minnesota and University of Virginia.

\end{document}